\documentclass{emulateapj}
\usepackage[breaklinks,colorlinks,citecolor=blue,linkcolor=magenta]{hyperref}
\usepackage{graphicx, epsf}
\usepackage{natbib}
\usepackage{color}
\usepackage{multirow}
\usepackage{appendix}

\newcommand{\Msun}{$M_\odot$}
\newcommand{\Mbh}{$M_{\rm BH}$}

\renewcommand{\deg}{\ensuremath{^{\circ}}}
\newcommand{\ml}{\emph{M/L}}
\newcommand{\mleff}{\emph{M/L}$_{\rm eff}$}
\newcommand{\mlpop}{\emph{M/L}$_{\rm pop}$}
\newcommand{\hst}{\emph{HST}}
\newcommand{\FI}{F814W}
\newcommand{\FB}{F336W}
\newcommand{\FV}{F547M}

\shortauthors{NGUYEN ET AL.}
 
\begin{document}


\title{Improved dynamical constraints on the mass of the central black hole in NGC~404}
\author{\mbox{Dieu D. Nguyen\altaffilmark{1}}}
\author{\mbox{Anil C. Seth\altaffilmark{1}}}
\author{\mbox{Mark den Brok\altaffilmark{1}}}
\author{\mbox{Nadine Neumayer\altaffilmark{2}}}
\author{\mbox{Michele Cappellari\altaffilmark{3}}} 
\author{\mbox{Aaron J. Barth\altaffilmark{4}}}
\author{\mbox{Nelson Caldwell\altaffilmark{5}}}
\author{\mbox{Benjamin F. Williams\altaffilmark{6}}}
\author{\mbox{Breanna Binder\altaffilmark{6}}}

\altaffiltext{1}{Department of Physics and Astronomy, University of Utah, 115 South 1400 East, Salt Lake City, UT 84112, USA\\
dieu.nguyen@utah.edu, aseth@astro.utah.edu, denbrok@physics.utah.edu}
\altaffiltext{2}{Max Planck Institute for Astronomy (MPIA), K\"onigstuhl 17, 69121 Heidelberg, Germany\\
~neumayer@mpia.de}
\altaffiltext{3}{Sub-department of Astrophysics, Department of Physics, University of Oxford, Denys Wilkinson Building, Keble Road, Oxford OX1 3RH, UK\\
michele.cappellari@physics.ox.ac.uk}
\altaffiltext{4}{Department of Physics and Astronomy, University of California, Irvine, Irvine, CA 92697-4575, USA\\
barth@uci.edu}
\altaffiltext{5}{Harvard-Smithsonian Center for Astrophysics, Harvard University, 60 Garden St, Cambridge, MA 02138, USA\\
caldwell@cfa.harvard.edu}
\altaffiltext{6}{Astronomy Department, University of Washington, Seattle WA 98195-1580, USA\\
ben@astro.washington.edu}
\altaffiltext{7}{Department of Physics and Astronomy, California State Polytechnic University, 3801 West Temple Ave, Pomona, CA 91768, USA\\
babinder@cpp.edu.}
\begin{abstract} 

We explore the nucleus of the nearby 10$^9$\Msun~early-type galaxy (ETGs), NGC~404, using \emph{Hubble Space Telescope} (\hst)/STIS spectroscopy and WFC3 imaging.  We first present evidence for nuclear variability in UV, optical, and infrared filters over a time period of 15~years.  This variability adds to the already substantial evidence for an accreting black hole at the center of NGC~404.  We then redetermine the dynamical black hole mass in NGC~404 including modeling of the nuclear stellar populations.  We combine \hst/STIS spectroscopy with WFC3 images to create a local color--\ml~relation derived from stellar population modeling of the STIS data.  We then use this to create a mass model for the nuclear region. We use Jeans modeling to fit this mass model to adaptive optics (AO) stellar kinematic observations from Gemini/NIFS.  From our stellar dynamical modeling, we find a  3$\sigma$ upper limit on the black hole mass of $1.5\times10^5M_{\odot}$.  Given the accretion evidence for a black hole, this upper limit makes NGC~404 the lowest mass central black hole with dynamical mass constraints.  We find that the kinematics of H$_2$ emission line gas show evidence for non-gravitational motions preventing the use of gas dynamical modeling to constrain the black hole mass.  Our stellar population modeling also reveals that the central, counter-rotating region of the nuclear cluster is dominated by $\sim$1~Gyr old populations.

\end{abstract}
\maketitle
\section{Introduction}\label{sec:intro}

Central massive black holes (MBHs) with masses of $\sim$10$^{6-9}$\Msun~are ubiquitous components of massive galaxies \citep[\textgreater$10^{10}$\Msun,][]{Kormendy13}. However, the demographics of black holes (BHs) in the centers of lower mass galaxies are less well constrained.  Any MBHs that exist in these low-mass (\textless$10^{10}$\Msun) galaxies are likely to have lower black hole masses, $\lesssim$$10^6$\Msun~\citep[e.g.,][]{Vandermarel04review,Greene07,McKernan11,Dong12,Neumayer12,Reines13,Yuan14}.  The presence of MBHs in lower mass galaxies (known as the occupation fraction) and the mass of these MBHs are sensitive to the mechanism that forms seed MBHs in the early universe  \citep{Volonteri10,vanWassenhove10}.  In fact, measurements of the mass and occupation fraction of MBHs in low mass galaxies are the only tools we currently have for constraining the formation of MBHs.

Unfortunately, the occupation fraction and masses of MBHs in low mass galaxies are difficult to measure because most galaxies are too far away to measure the dynamical effect of an $\lesssim$10$^6$\Msun~MBH with current instrumentation.  Thus, our knowledge of  MBHs in low mass galaxies are limited to galaxies where active galactic nuclei (AGN) are observed \citep[e.g.,][]{Filippenko89,Greene04,Barth04,Satyapal07,Reines13,Moran14,Baldassare15}, or where tidal disruption events (TDEs) occur \citep[e.g.,][]{Maksym14}.  These objects provide strong evidence that some MBHs do exist in galaxies with stellar mass as low as a $\sim$3$\times10^8$\Msun.  However, evidence for an AGN is detected in at most a few percent of galaxies \citep{Moran14}, while rough MBH mass estimates based on broad line AGN are available for only a small number of objects \citep{Greene07,Reines13,Reines15,Baldassare15}.  Dynamical estimates are only possible in the very nearest galaxies, and only a handful of dynamical MBH mass estimates or significant non-detections are available  \citep{Gebhardt01,Verolme02,Valluri05,Seth10,Denbrok15}.  This lack of good MBH mass estimates also means that scaling relations between galaxy properties (e.g., bulge luminosity and dispersion) lack data at the low-mass end \citep[see recent compilations][]{Kormendy13,McConnell13b,Graham15,Saglia16}.  Additional dynamical BH mass measurements in low-mass galaxies are therefore important for understanding the physics that underlies known scaling relations.

Most low-mass galaxies also host massive nuclear star clusters  \citep[NSCs, e.g.,][]{Boker02, Cote06,Denbrok14,Georgiev14}.  There are a number of galaxies which contain both an MBH and an NSC \citep[e.g.,][]{Filippenko03,Seth08a}.  However, the relationship between MBHs and NSCs is not yet well understood.  Initial discoveries of scaling relationships between NSCs and ETGs suggested that low-mass galaxies may form NSCs rather than MBHs \citep{Ferrarese06,Wehner06}, more recent studies suggest a transition with galaxies increasingly favoring MBH in galaxies of higher mass \citep{Graham09,Neumayer12,Kormendy13,Georgiev16}.  This transition may occur due to tidal disruption effects \citep{Antonini13, Antonini15a, Antonini15b} or feedback \citep{McLaughlin15,Nayakshin09b}.

NSCs can provide a mechanism for forming seed MBHs \citep[e.g.,][]{portegieszwart04b} or grow concurrently with MBHs \citep[e.g.,][]{Hopkins09}.  However, the existence of an MBH around which an NSC appears to be currently forming in the nearby galaxy Henize 2-10 suggests that MBHs can form independently of NSCs \citep{Nguyen14}.   Several galaxies containing both an NSC and an MBH candidate are currently known. These include the Milky Way \citep{Genzel10, Schodel14}, M31 \citep{Bender05}, NGC~4395 \citep{Filippenko03, Denbrok15}, NGC~1042 \citep{Shields08}, NGC~3621 \citep{Barth09}, NGC~4178 \citep{Satyapal09}, and NGC~3367 and NGC~4536 \citep{McAlpine11}.

The nearby galaxy NGC~404 provides an excellent opportunity to study the relationships of NSCs and MBHs.  NGC~404 is a dwarf S0 galaxy \citep[$M_{\star}\sim10^9$\Msun,][]{Seth10} at a distance of just $3.06\pm0.37$~Mpc \citep{Karachentsev02}.  The galaxy appears to host an MBH within its prominent central NSC \citep{Seth10, Binder11, Nyland12, Paragi14}.

The NSC of NGC~404 was studied in depth by \citet[][hereafter S10]{Seth10}.  They found a dynamical mass of the NSC of ($1.1\pm 0.2$)$\times10^7$\Msun, and stellar populations analysis suggest that half this mass formed $\sim$1~Gyr ago.  Gas and star formation in the outer parts of the galaxy suggest that NGC~404 acquired gas $\sim$1~Gyr ago during a minor gas-rich merger \citep{delrio04, Thilker10, Bouchard10}; S10 proposed the $\lesssim$1~Gyr stars in the NSC were formed due to this merger.   However, consistent with its early-type morphological classification, the galaxy is very old, with 90\% of the stellar mass being formed more than 8 Gyr ago \citep{Williams.B.F10}.

Recent observations have shown that the nucleus of NGC~404 appears to host an accreting BH that powers a low-luminosity AGN.  The first indication was from the optical spectrum, which was found to have a LINER classification \citep{Stauffer82, Keel83, Ho97}; there is considerable debate on the nature of LINER nuclei \citep[e.g.,][]{Singh13}, but they have frequently been found to be associated with AGN \citep[e.g.,][]{Nagar05}.  The nucleus has also been found to have a hard and compact X-ray core with $L_X=1.3^{+0.8}_{-0.5}\times10^{37}$ erg s$^{-1}$ \citep{Binder11}, and a compact radio core \citep{Nyland12} at arc-second scales.  \citet{Maoz05} also found the UV emission is variable, declining by a factor of $\sim$3 between 1993 and 2002 while S10 also find some evidence for variability in the NIR due to hot dust around the AGN.  The mid-IR spectrum of NGC~404 shows evidence for high excitation consistent with other AGNs \citep{Satyapal04}.  In particular, the ratio of the [\ion{O}{4}] flux relative to other emission lines ([\ion{Ne}{2}], [\ion{Si}{2}]) is higher than any other LINERs in the sample of \citet{Satyapal04} and is similar to other known AGNs. However, [\ion{Ne}{5}] lines, which are the more reliable indicator of AGN activity \citep{Abel08}, are not detected.  Among these, we find the presence of a hard X-ray core and the variable UV emission to be the strongest pieces of evidence for an AGN in NGC~404.

Despite this evidence, the proof of an AGN in NGC~404 is complicated by recent star formation in the nucleus.  \hst~observations of H${\alpha}$ shows a compact emission source at 0$\farcs$16 north of the nucleus that may be due to supernova remnants \citep{Pogge00}.  The nuclear UV spectrum also clearly shows evidence for massive stars although these observations leave room for a UV continuum component from an AGN component \citep{Maoz98}.  \citet{Paragi14} did not detect the radio source on 10~mas scales, suggesting either a non-active or low mass BH, and suggesting that at least some of the radio emission is due to star formation. An upcoming paper by Nyland et al. {\em (in prep)} bridges the scale of emission sensitivity, and does indeed find some compact radio emission coincident with the nucleus.

Dynamical measurements using high-resolution AO data from Gemini/NIFS by S10 indicate the possible detection of an MBH, with a firm upper limit of \textless$10^6$\Msun.  More specifically, S10 used two kinematic tracers to constrain the BH mass, stellar kinematics from the CO band-head at 2.3$\mu$m and gas kinematics from the H$_2$ line at 2.12$\mu$m.  Their stellar dynamical modeling gave a best fit \Mbh~= $5\times10^4$\Msun, but consistent with 0~\Msun~at 3$\sigma$.  The kinematic dynamical model of molecular hydrogen emission from the nucleus revealed the best-fit mass for the BH in a large range of a \Mbh~= $4.5^{+3.5}_{-2.0}\times10^5$\Msun~(3$\sigma$ error bars).   Although there is a large discrepancy between the BH mass estimate from the stellar and gas modeling, S10 put a firm upper limit of \Mbh$<10^6$\Msun.

The dynamical modeling in S10 depended on a number assumptions, most importantly the assumption of a constant \ml~throughout the nucleus.  This is a poor assumption due to the obvious spatial gradients of stellar populations within the nucleus.  The stellar mass in the central resolution element of the Gemini/NIFS data was comparable to the BH mass (i.e.,~the sphere of influence was not well resolved), hence, any spatial gradients in the \ml~can have a significant effect on the BH mass estimate.  In this paper, we use new STIS spectroscopy and WFC3 imaging to quantify the spatial variations in the \ml~throughout the NGC~404 nucleus and improve the BH mass estimate using the same Gemini/NIFS data presented in S10.  Incorporating \ml~gradients to refine BH mass estimates was recently explored by \citet{McConnell13a}, who used color gradients to explore possible radial \ml~gradients in three giant elliptical galaxies.

This paper is organized into seven sections. In Section~\ref{sec:data} we describe the observations and discuss their initial reduction and analysis.     The nuclear variability, stellar population modeling of the STIS data, and a creation of color--\ml~and mass maps are presented in Section~\ref{sec:m2lvary}.   Jeans modeling of the stellar kinematics incorporating the mass map is presented in  Section~\ref{sec:jeans}, and the gas dynamical modeling is in Section~\ref{sec:gasmodeling}. We discuss and conclude in Section~\ref{sec:dis} and Section~\ref{sec:concl}.  We assume a distance of $3.06\pm0.37$~Mpc \citep{Karachentsev02}, giving a physical scale of 15~pc arcsec$^{-1}$. Unless otherwise indicated, all quantities quoted in this paper have been corrected for a foreground extinction $A_V=0.306$ \citep{Schlafly11} using the interstellar extinction law of \citet{Cardelli89}.

\section{Data}\label{sec:data}
\subsection{HST/STIS Spectroscopy}\label{ssec:stis} 

The STIS spectroscopic observations (PID: 12611, PI: Seth) were taken on 2012 November 18 with the G430L grating and the 52$\farcs$0 $\times$ 0$\farcs$1 slit.  This provides spectra over a wavelength range of 2,900~\AA~to 5,700~\AA~with a pixel size of 2.73~\AA, and spectral resolving power of R $\sim$ 530--1,040.  Seven individual exposures were taken, each with an exposure time of $\sim$1890~s, for a total exposure time of 13230~s.  The source was dithered along the slit and centered near the E1 aperture position to minimize charge-transfer inefficiency losses.

Reduced and rectified spectroscopic exposures ({\tt x2d} files) were downloaded from the Hubble Legacy Archive (HLA).  The STIS images suffer from numerous imaging defects, especially vertical defects extending over many rows.   To remove these features from our dithered data, we created a median image without dithering which was subtracted from each individual, dithered image.   Each dither was separated by nine pixels, thus, some background galaxy light will be included in this median image.  However, this effect is very small; the maximum row-averaged flux of the median image is \textless 25\% of that at the outermost radius we analyze.

We then combined the seven individual exposures into a single two-dimensional (2--D) spectrum.  We applied integer offsets of nine pixels between each exposure after verifying these offsets using the position of the nucleus. Before combining exposures, we used sigma-clipping at the 3$\sigma$ level to remove cosmic rays and bad pixels, and also flag bad pixels based on their data quality (DQ) values.  Specifically, we included pixels with DQ~\textless~1000.  We then combined the exposures by taking a mean value of all remaining unflagged and unclipped pixels from the seven exposure stack.  We also created an error frame for the combined image by creating variance frames from the original error frames.  Then for each pixel, we added the appropriate variance frames together and divided by the square of the total number of exposures.

The signal-to-noise (S/N) of the central pixel in the combined spectrum is 37 and 50 per pixel at 3700 and 5000~\AA, respectively. The S/N drops off to 20 per pixel at 5000~\AA~for binned data at +(0$\farcs$475--0$\farcs$575) and -(0$\farcs$675--0$\farcs$825), which are the outermost data we use in our spectral analysis.

\subsection{HST/WFC3 $\&$ WFPC2 Data}\label{ssec:hst}   
 The \hst~imaging data we use in this work are taken from different cameras. They include WFC3 $\&$ WFPC2, ACS/WFC, $\&$ ACS/HRC imaging; the data we use in this work are summarized in Table~\ref{tab_data}. Reduced drizzled images were downloaded from the \hst/HLA.  The \hst/WFC3 images were obtained contemporaneously with the above STIS spectroscopy observation, and we use these in combination with the spectroscopic data to create a 2--D mass map.  We discuss the creation of point-spread functions for the \hst~data in Section~\ref{ssec:psf}. 

\begin{table*}[ht]
\caption{\hst/WFC3 $\&$ WFPC2 Data}
\centering
\begin{tabular}{ccccccc}
\hline \hline 
Filter                &  Camera  &  Aperture     &  UT Date &Exposure Time &  Plate Scale     &   PID \\[0.1cm]  
                        &                 &                    &                 &       [s]              &[$\arcsec$/pixel]&         \\[0.1cm]    
\hline
                        &                 &                    & WFC3     &                         &                          &         \\[0.1cm] 
\hline 
F160W            & WFC3/IR  &  IRSUB256     &2012-11-18&6$\times$3.33 &  0.04           & 12611 \\[0.1cm] 
\FB                   & WFC3/UVIS&UVIS2-C512C-SUB&2012-11-18&4$\times$200  &  0.04           & 12611 \\[0.1cm] 
F502N                 & WFC3/UVIS&UVIS2-C512C-SUB&2012-11-18&2$\times$110  &  0.04           & 12611 \\[0.1cm] 
\FV                   & WFC3/UVIS&UVIS2-C512C-SUB&2012-11-18&4$\times$100  &  0.04           & 12611 \\[0.1cm] 
\FI                   & WFC3/UVIS&UVIS2-C512C-SUB&2012-11-18&4$\times$50   &  0.04           & 12611 \\[0.1cm]  
\hline
                      &          &               &WFPC2 PC \& ACS HRC   &      &             &       \\[0.1cm] 
\hline    
\FB                   & ACS/HRC  &               &2002-10-28&2$\times$300  &  0.025          &  9454 \\[0.1cm] 

\multirow{2}{*}{F555W}&   WFPC2  &  PC1          &1995-09-06&2$\times$320  &  0.05           &  5999 \\[0.1cm]        
                      &   WFPC2  & PC1-FIX       &1996-10-21&2$\times$350  &  0.05           &  6871 \\[0.1cm] 
\FI                   &   WFPC2  & PC1           &1995-09-06&2$\times$300  &  0.05           &  5999 \\[0.1cm] 
\hline
\end{tabular}
\tablenotemark{}
\label{tab_data}
\end{table*}

\subsection{Gemini NIFS Data}\label{ssec:nifs} 

NGC 404 was observed with Gemini/NIFS on 2008 September 21-22 using the Altair laser guide star system. The nucleus was observed for a total of 4560~s (see S10 for details).  These data were used to derive stellar kinematics (from the CO band-head region) and the molecular hydrogen kinematics (from the 2.12 $\mu$m H$_2$ 1-0S(1) emission line) by S10; we use these data in the stellar and gas dynamical modeling presented in this work in Section~\ref{sec:jeans} and Section~\ref{sec:gasmodeling}, respectively.

We briefly review the derivation of the stellar kinematics here; more details are found in S10 and \citet{Seth08b}. We use the penalized pixel fitting (pPXF) code of \citet{Cappellari04} to measure the stellar line-of-sight velocity distribution (LOSVD) including the radial velocity, velocity dispersion,  and the non-Gaussian terms \emph{h}$_3$ and \emph{h}$_4$, which measure the skewness and kurtosis of the LOSVD \citep{vanderMarel93}.  We fit the strong CO bandheads in the wavelength range from 22,850~\AA~to 23,900~\AA. Kinematics were derived using eight spectra from the templates of \citet{Wallace96}; these include spectral types from G to M (which show significant CO absorption) and classes from the main sequence to supergiants.  We measured the instrumental resolution of each spatial pixel using sky lines.  The LOSVD errors were estimated via Monte Carlo simulations using a propagated noise spectrum.  Errors in the radial velocities ranged from 0.5 to 6~km s$^{-1}$ changing with S/N (see Figure 8 of S10).

One feature of this data set that is important in our analysis is the assumed center of the NGC~404 NSC.  S10 used both a dust-emission corrected photocenter at R.A. = 01$^{\rm h}$09$^{\rm m}$27$\fs$02 and Decl. = 35\deg43$^{\prime}$5$\farcs$1) for modeling the gas kinematics and a kinematic center at R.A. = 01$^{\rm h}$09$^{\rm m}$27$\fs$01 and Decl. = 35\deg43$^{\prime}$4$\farcs$8 for modeling the stellar kinematics.  The stellar kinematic center is about half a pixel ($\sim$0$\farcs$025) south of the photocenter. The uncertainties on these measurements are $\sim$0$\farcs$02.  We will show in Section~\ref{sec:gasmodeling} that this uncertainty in center translates into large variations in our gas dynamical BH mass estimates, suggesting the gas dynamics results are not robust.

\begin{figure}[ht]
     \centering
      \epsscale{1.2}
          \plotone{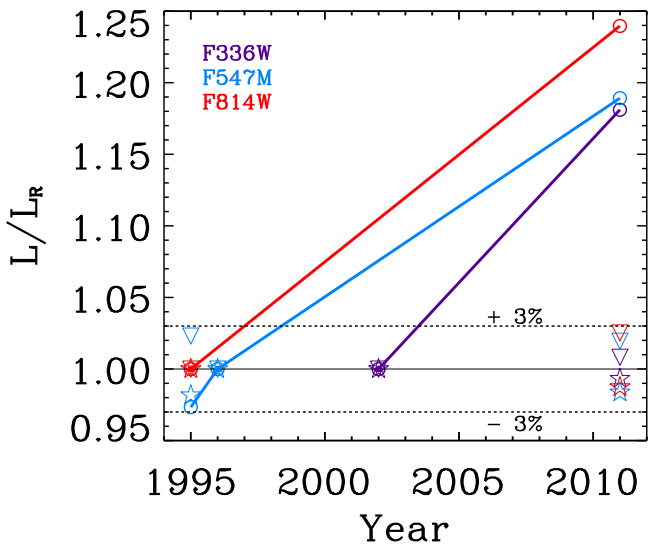}
          \caption[SCP06C1]{Evidence of nuclear variability in several filters over a 15 year period. Open purple, blue, and red circles represent nuclear luminosities (\textless$0\farcs2$) in the \FB, \FV, and \FI~filters; each luminosity is presented as a ratio relative to the epoch shown with filled circles (L$_{\rm R}$).  The upside-down triangles and five-pointed star are the ratios of luminosities in two annuli surrounding the nucleus with radii of (0$\farcs$2--0$\farcs$5) and (0$\farcs$5--1$\farcs$0) respectively; these show much less variation than the nuclear flux, showing that differences between filter responses between different cameras and sky subtraction issues affect the measurements below the 3\% level.}
\label{nucleus_vary}
\end{figure}

\subsection{PSF Determination} \label{ssec:psf} 

The different point-spread functions (PSFs) for our \hst~images at different filters and our Gemini/NIFS spectra is a complicating factor in this analysis.  The \hst~PSFs play a significant role in creating the mass map of the nucleus while the Gemini/NIFS PSF is critical for the dynamical modeling. In this work, we use two different types of PSFs: (1) for the \hst/WFC3 data, we use {\tt Tiny Tim} PSFs \citep{Krist95, Krist11}, and (2) for the Gemini/NIFS data we use a two-component inner Gaussian + outer Moffat profile PSF presented in S10.

For the \hst/WFC3 PSFs, we create the model PSF for each WFC3 exposure using the {\tt Tiny Tim} routine for each of the four individual {\tt flt} exposures.  The PSFs were created using the $\tt{tiny3}$ task, which includes a charge diffusion kernel and the distortion of a PSF.   We then insert these into mock {\tt flt} images at the position of the nucleus in each individual exposure to simulate our observations.  Finally, for each filter, we combine the four PSFs at each of the four dither positions into the final PSF using the {\tt Astrodrizzle} package \citep{Avila12}.

For the Gemini/NIFS images, we use the two component PSF fit to the NIFS data from S10.  The fits are made by convolving an \hst~image of the NGC~404 nucleus to match the NIFS continuum image using a two component PSF: (1) an inner Gaussian component, and (2) a larger scale Moffat profile $(\Sigma(r)=\Sigma_0/[(1+(r/r_d)^2)]^{4.765})$) whose size is constrained by fits to the telluric calibrators taken alongside the science images. S10 find the best fit model PSF with an inner Gaussian FWHM of $\sim$0$\farcs$12, and an outer Moffat with FWHM of $\sim$0$\farcs$95 profiles; each containing about half of the light.

\begin{figure*}[!htb]
\minipage{0.5\textwidth}
 \includegraphics[width=\linewidth]{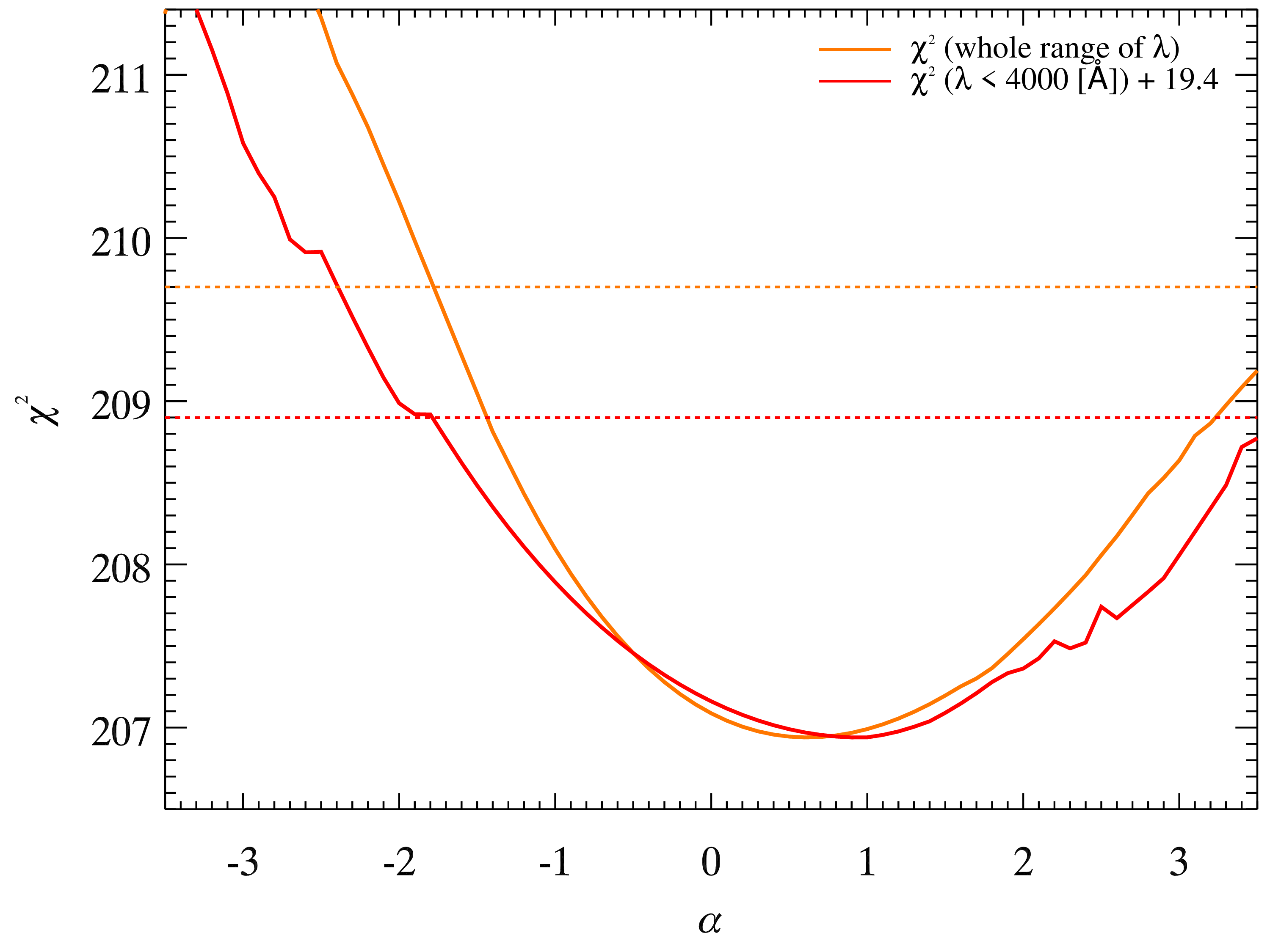}\label{agn_index}  
\endminipage\hfill
\minipage{0.5\textwidth}
  \includegraphics[width=\linewidth]{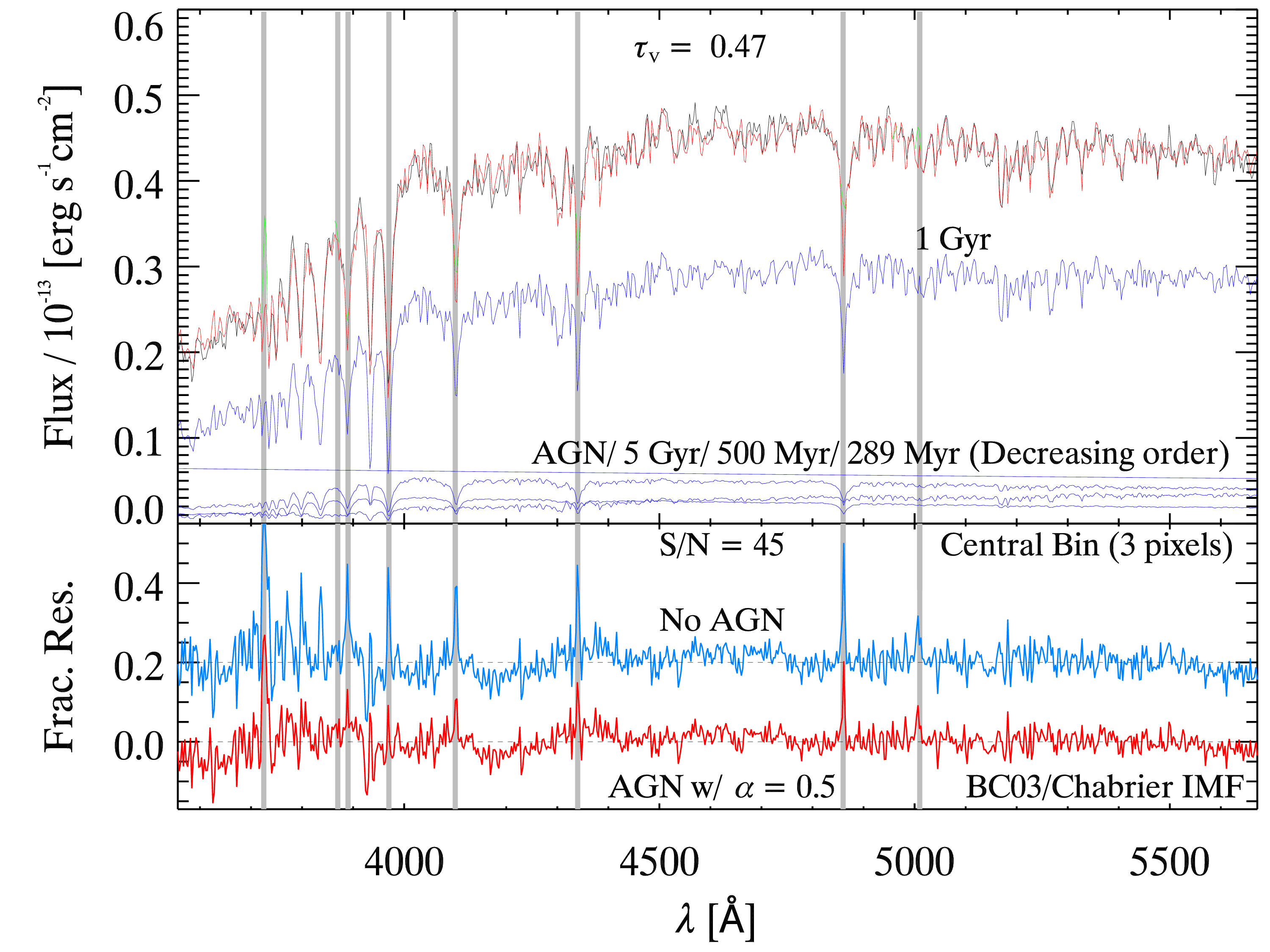}\label{agn_vs_noagn}
\endminipage
\caption[SCP06C1]{Left panel: $\chi^2$ values to determine the best-fitting spectral index for the AGN component in our stellar population synthesis fits.   The orange line is $\chi^2$ over the whole wavelength range, while the red line shows just the $\chi^2$ in the blue regime (\textless4,000~\AA) where the AGN is likely to be most prominent. The dashed horizontal lines show the best-fit $\chi^2$ without an AGN component included in the fit over both wavelength ranges.  The $\chi^2$ of the fits taken only below 4,000~\AA~have had a value of 19.4 added to them.  Right upper panel:  The central STIS spectrum of NGC~404 (summed over three spatial pixels) is shown in black, the best-fitting stellar population synthesis model fits \emph{including an AGN component} with $\alpha=0.5$ shown in red.  Vertical grey lines show emission line regions that were excluded from the fit. Blue spectra indicate the different ages SSPs and AGN component contributing to the multi-age fit.   Right bottom panel: fractional residuals of the best-fit multi-age  spectrum with the best-fit AGN component (red line) and without an AGN component (blue line). The no AGN fit residuals are offset by $+$0.2.}
\label{agn_ssp}
\end{figure*} 

\subsection{Astrometry}\label{ssec:astrometry} 
 
Astrometry is an essential step in this work.  Below we outline the astrometric alignments we carried out:
\begin{enumerate}

\item Due to the small field of view of our WFC3 data, we obtain absolute astrometry for the images by aligning an ACS \FI~image in which the NSC is saturated to the 2MASS point source catalog.  The alignment uses 12 unsaturated sources with a root mean squared offset of 0$\farcs$20; this represents our absolute astrometric accuracy.  We then align our WFC3 \FI~image to this image using bright compact stars around the core.

\item Astrometric alignment of NIFS, WFPC2, ACS/HRC, and the new WFC3 data were then tied to the WFC3 \FI~data.  The centroid position of the nucleus in WFC3 \FI~was used as the reference position to align up each image.   Although dust clearly affects the area around the nucleus, it appears that the center of the nucleus does not suffer from significant internal extinction as seen in the \FV--\FI~color map (Figure~2 of S10) and the \FB--\FI~color map (this work, Figure~\ref{colormap}).  The alignment on the nucleus can be verified by comparing the alignment of sources in these color maps with as well as the alignment of the NIFS Br${\gamma}$ map and the \hst~H${\alpha}$ image (Figure~6 in S10).

\item Finally, we align the STIS spectroscopy to the WFC3 image. To do this, we collapse the spectra to create a 1--D image of the slit, multiplying by the appropriate filter curve to make synthetic \FV~and \FB~images. We then compare these 1--D images to the astrometrically aligned WFC3 images in the same filters to obtain WCS information on the slit positions.  Because of the low S/N in the STIS image at large radius, the fitting of the 1--D images was performed at radii \textless0$\farcs$8.  Both image comparisons gave consistent WCS coordinates. 

\end{enumerate} 

\section{Mass-to-light ratio variations}\label{sec:m2lvary}
\begin{figure*}[!htb]
\minipage{0.5\textwidth}
  \includegraphics[width=\linewidth]{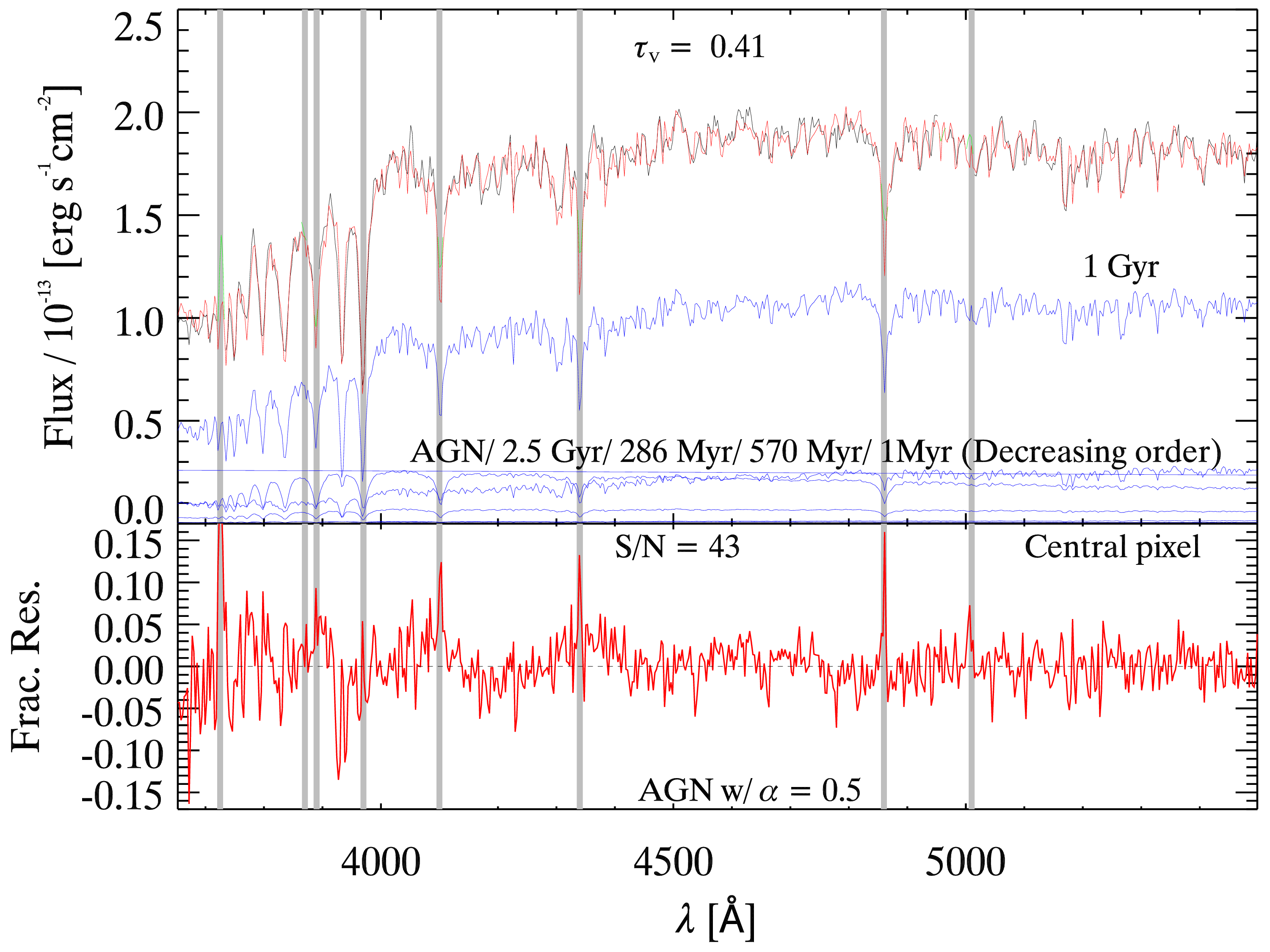}\label{ssp_agn_fit}
\endminipage\hfill
\minipage{0.5\textwidth}
  \includegraphics[width=\linewidth]{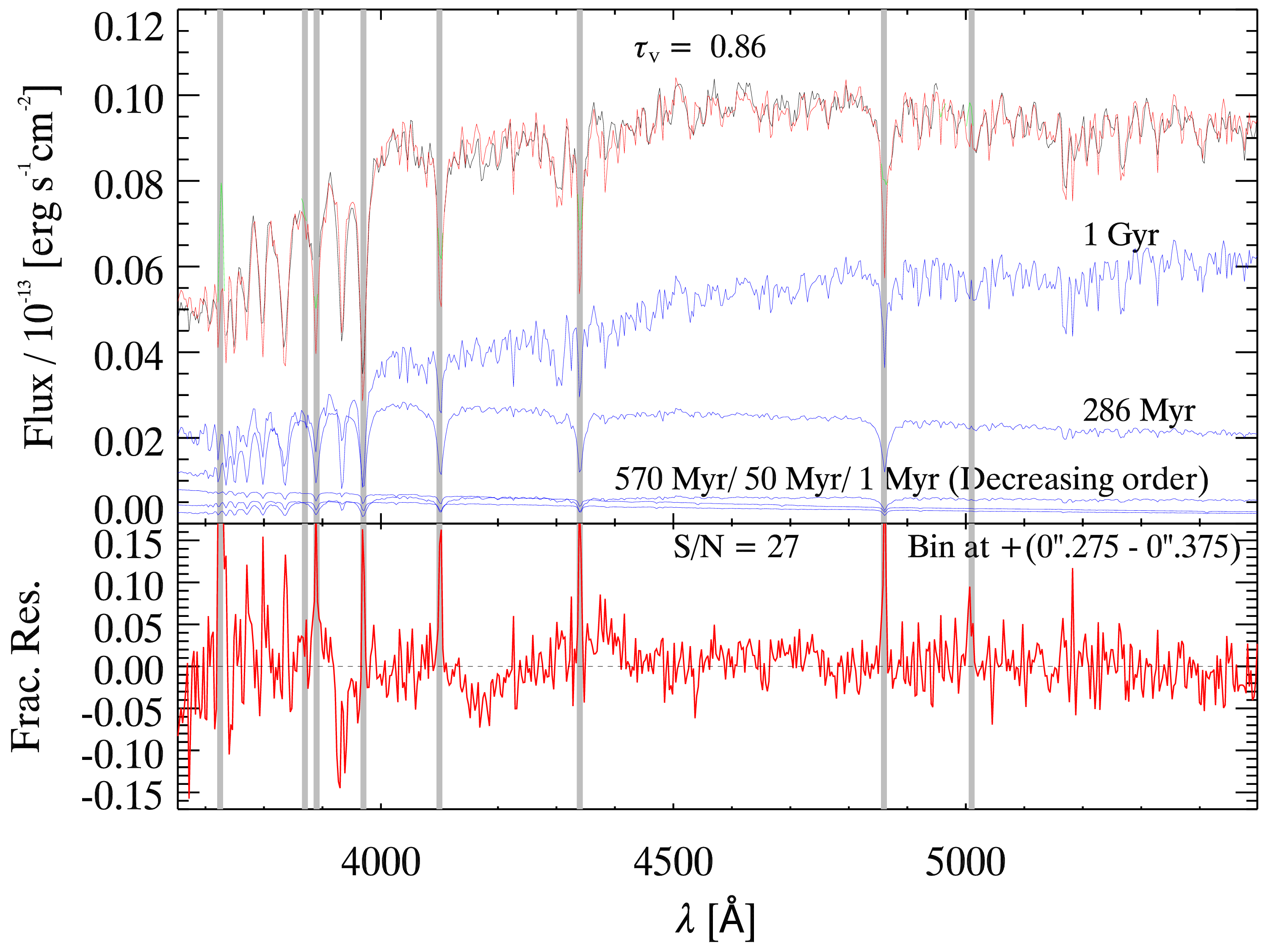}\label{ssp_noagn_fit}
\endminipage
\caption[SCP06C1]{Left panel: best-fit stellar population models to individual pixel spectra at the center. Right panel: best-fit stellar population models to a binned spectrum at large radius spanning radii of $+($0\farcs275--0\farcs375).  The central spectrum  is fitted with an AGN continuum component, while the outer bin is not.  Markings otherwise as in the right panel of Figure~\ref{agn_ssp}.}
\label{ssp_fit}
\end{figure*}
\begin{table*}[ht]
\caption{Stellar population fits to the STIS central pixel}
\centering
\begin{tabular}{cccccccccccccc}
\hline \hline 
Model                    &Age [Myr]& AGN   & 1     & 10    & 50    & 100   & 286   & 570   & 1000  & 2500  & 5000  & 10,000 & 13,000\\[0.1cm]
                         &Z        &       & 0.02  & 0.02  & 0.02  & 0.02  & 0.02  & 0.02  & 0.02  & 0.02  & 0.008 & 0.008  & 0.0001\\[0.1cm]
\hline    
\multirow{2}{*}{AGN}     &    l    & 0.170 &  --   &   --  &   --  &  --   & 0.026 & 0.029 & 0.634 & 0.152 &  --   &   --   &  --   \\[0.1cm]  
                         &    m    &  --   &  --   &   --  &   --  &  --   & 0.028 & 0.034 & 0.735 & 0.204 &  --   &   --   &  --   \\[0.1cm]
\hline     
\multirow{2}{*}{No AGN}  &    l    &  --   &  --   & 0.031 &   0   & 0.108 &  --   &   --  & 0.743 & 0.118 &  --   &    --  &  --   \\[0.1cm] 
                         &    m    &  --   & --    & 0.002 &   0   & 0.024 &  --   &   --  & 0.707 & 0.267 &  --   &    --  &  --   \\[0.1cm]
\hline 
\end{tabular}
\tablecomments{Best-fit stellar population of the central bin of the STIS spectrum.  The upper model includes an AGN continuum component with power-law slope $\alpha=0.5$, while the lower panel does not include an AGN component. The best-fit light and mass fractions are given for each component (AGN or BC03 SSP) in the model. Here, l (light fraction) and m (mass fraction) for each best-fit model.}
\label{tab_pop}
\end{table*}

\subsection{Constraining the AGN Continuum Contribution Through Variability}\label{ssec:nucleusvary}

In this section, we describe the procedure to characterize the possible AGN continuum contribution to the nuclear spectrum through examining the time variability of the nucleus in different bands.  Nuclear variability has been suggested by several previous observations \citep{Maoz05,Seth10}, but here we provide the strongest evidence yet for broad-band nuclear variability in NGC~404.  This variability is important both in providing additional evidence for an accreting BH in NGC~404, and for constraining our stellar \ml~modeling.

The most compelling measurements to date have been the UV variability presented by \citet{Maoz05}.  Comparing their 2002 \hst/ACS HRC UV observations to 1994 \hst/FOS spectra with an $0\farcs86$ aperture \citep{Maoz98} and 1993 pre-correction \hst/FOC imaging \citep{Maoz05}, they suggest that the nuclear UV flux at 2500~\AA~in 2002 was a factor of $\sim$3 smaller than in the 1993-94 observations; this drop is consistent with the possible AGN continuum visible in their FOS spectra.  Because that work had a number of similar datasets where fading was not seen, they argue that this drop in flux is robust despite the very different data sets and large apertures used in the comparison.  Variable hot dust emission in the NIR was suggested by S10 based on the comparison of the 2008 NIFS $K$-band images with 1998 \hst/NICMOS NIC2 observations in F160W.  Within the central 0$\farcs$2, the NIFS $K$-band image is 30--40\% fainter than the NIC2 observations.  This picture is consistent with the observed spectral signature of hot dust emission in the nucleus, but the comparison across bands and instruments leads to significant uncertainties (S10).  The observations we present below provide more robust evidence for nuclear variability than any previous sets of observations.

We examine the variability of the nucleus using the new WFC3 images (all taken 2012-11-18) along with images in similar filters from WFPC2 and ACS/HRC (Table~\ref{tab_data}).  There are repeat measurements in three filters: (1) our WFC3 \FB~image can be compared to a previous ACS/HRC F330W image taken 2002-10-28 \citep{Maoz05}, (2) our WFC3 \FV~image can be compared to previous WFPC2 images in F555W~(1996-10-21) and F555W (1995-09-06), and (3) our WFC3 \FI~image can be compared to a previous WFPC2 \FI~image (1995-09-06).

To eliminate mismatches in the background level, we first estimated the sky fluxes in an annulus of 10$\farcs$0--13$\farcs$0 away from the nucleus (the maximum radius available in all observations).  These sky levels were then subtracted off of all nuclear photometric measurements. Next, we performed aperture photometry in three radial ranges: the nucleus (\textless0$\farcs$2), and two annuli surrounding the nucleus, from 0$\farcs$2--0$\farcs$5 and 0$\farcs$5--1$\farcs$0.  The photometry was converted into luminosities using appropriate zero-points for each observation.

We compute the ratio of luminosities between two epochs to gauge the level of nuclear variability.  Figure~\ref{nucleus_vary} shows the nuclear luminosities (\textless0$\farcs$2) increase by 18--25\% in all three bands between 1994--2012. Specifically, nuclear luminosities increase $\sim$18\% for \FB, $\sim$25\% for F555W (\FV), and $\sim$24\% for \FI~band.  For each filter, we choose one epoch to be the reference epoch (filled circles in Figure~\ref{nucleus_vary}) and compare the other epochs to the reference epoch.  We use a 0$\farcs$2 aperture to capture a majority of the nuclear flux; the encircled energy within 0$\farcs$2 is 90\%, 87\%, and 85\% for the WFC3 \FB, \FV, and \FI~filters, respectively. The outer two annuli (triangles and plus signs in Figure~\ref{nucleus_vary}) can then be used to gauge the variations that might be caused due to bandpass and detector sensitivity.  We find variations of \textless3\% at all epochs for these two annuli; we note that formal uncertainties are much smaller than this.  Thus, the nuclear variations we find are highly significant (6-8 times the largest variations seen in our annuli), providing the strongest evidence to date for broadband UV-NIR variability in the NGC~404 nucleus.

The increase in nuclear UV (\FB) flux we see from 2002--2012 is opposite to the decrease in 2500~\AA~UV flux observed by \citet{Maoz05} from 1993--2002; this argues against a single transient event (e.g., supernova) being responsible for the variability.  Given that the increase in flux at all wavelengths is opposite to apparent decrease in the NIR observed by S10 over overlapping time periods suggests that the variability likely occurs on time scales smaller than the decade scales probed here.  Finally, we note that the STIS data analyzed below were taken contemporaneously with the WFC3 data.  The variability we see here is consistent with the $\sim$17\% contribution of AGN continuum at 3,700~\AA~that provides the best-fit to the nuclear spectra (Section~\ref{ssec:ssp}). 

\subsection{Stellar Population Synthesis Models of the Nucleus}\label{ssec:ssp} 

We fit a range of single stellar population (SSP) models to the nuclear STIS spectroscopic data to determine the ages of stars and~\ml~spatial variations within the NGC~404 NSC. We show that the nuclear spectrum is somewhat better fit with the addition of an AGN continuum component, and use this as our default model in the nuclear regions.  The stellar population of the NSC shows evidence of a range of ages, but an intermediate age component (1--5 Gyr), is particularly dominant in the central \textless0$\farcs$3 of the NSC.

\subsubsection{SSP Model Fitting Methodology}\label{sssec:model_setup}

We fit the nuclear STIS spectroscopy to two SSP models including (1) the \citet[][BC03]{Bruzual03} models, and (2) the updated version of the BC03 models incorporating the asymptotic giant branch (AGB) models \citep{Marigo07, Marigo08} obtained from St\'ephane Charlot, which we refer to as CB07 models. These models include spectra for single age simple stellar populations over a grid of ages and metallicities at a spectral resolution similar to our observations. As in S10, the ages used were 1, 10, 50, 100, 300, 600, 1,000, 2,500, 5,000, 10,000, 13,000 Myr.  In the absence of knowledge about the age-metallicity relation, we assume a relation similar to that of M54, the nucleus of the Sgr dwarf galaxy \citep{Siegel07};  the metallicity is assumed to be Solar while at the oldest age it drops to Z = 0.0004 ([M/H ] = -1.7).  We note that \citet{Bresolin13} find the present day metallicity in the outer disk of NGC~404 is $\sim$80\% Solar; given the BC03 models available and the likelihood of enhanced metallicity at the center of our galaxy, this justifies our use of Solar metallicity templates for younger ages.

We also correct for the poor wavelength calibration of the STELIB library used by BC03 and CB07, as identified by \citet{Koleva08}. To remove these issues, we compare 100~\AA~sections of the observed spectrum to the BC03 models to derive a velocity offset (relative to the systematic velocity of NGC 404). We find the velocity corrections to be $\pm$35 km s$^{-1}$.  We then fit the BC03 and CB07 models to our STIS spectroscopy using the Simplefit program \citep{Tremonti04}, which uses a Levenberg--Marquardt algorithm to find the best-fit parameters and scales of the input model spectra by performing a $\chi^2$ minimization. The program also calculates the extinction of the STIS spectroscopy by using the simple model for dust absorption using the prescription of \citet{Charlot00}. We exclude regions around expected emission lines from the stellar population fit.

To optimize our fit, we fit the flux-calibrated spectrum between 3,500--5,700~\AA.   We chose this wavelength range by examining the $\chi^2$ of the nuclear fit (with and without an AGN component, see below) as we change the blue end of the fit where the data become lower S/N.  We vary the blue end of the fitted wavelengths from 3,000 to 4,500~\AA~and find that the optimal blue extreme wavelength is at 3,500~\AA~for both BC03 and CB07 SSP models.  The best-fit for the BC03 templates had a reduced $\chi^2$ of 0.98 (670 degrees of freedom). 

\subsubsection{Nuclear Spectrum $\&$ Testing an AGN continuum}\label{sssec:agn_ssp}  

\begin{table*}[ht]
  \caption{Resolved Stellar Population Fitting Results}
\hspace{-0.3cm}
\begin{tabular}{ccccccccccccccccc} \hline \hline 
$r$                            &Frac.&   AGN  & 1     & 50    & 100   & 286   & 570   & 1.0   & 2.5   & 5.0   &$\tau_V$&$M/L_I$& $M/L_I$ &$\chi^2_{\rm r}$&$M/L_{I}^{\star}$\\
$['']$                         &     &        & [Myr] & [Myr] & [Myr] & [Myr] & [Myr] & [Gyr] & [Gyr] & [Gyr] & BC03   &  BC03 & CB07    &   BC03       & BC03 \\[0.1cm]  
     
   (1)                         & (2) &   (3)  & (4)   & (6)   & (7)   & (8)   & (9)   & (10)  & (11)  & (12)  & (13)   &  (14) &  (15)   &    (16)      & (17) \\[0.1cm] 
\hline   
\hline   
\multirow{2}{*}{+(0.475--0.575)} &  l  &$\times$&  --   &  --   &   --  & 0.034 & 0.046 & 0.920 & --    & --    &  0.98  &  0.85 &   0.89  &   1.71       & --   \\[0.1cm] 
                               &  m  &$\times$&  --   &  --   &   --  & 0.081 & 0.065 & 0.854 & --    & --    &  --    &  --   &   --    &   --         & --   \\[0.1cm]
\hline   
\multirow{2}{*}{+(0.375--0.475)} &  l  &$\times$& 0.030 & 0.025 &   --  & 0.283 & 0.058 & 0.602 & --    & --    &  0.86  &  0.52 &   0.51  &   1.25       &  --  \\[0.1cm] 
                               &  m  &$\times$& 0.002 & 0.038 &   --  & 0.257 & 0.093 & 0.615 & --    & --    &  --    &  --   &   --    &   --         &  --  \\[0.1cm]  
\hline
\multirow{2}{*}{+(0.275--0.375)} &  l  &$\times$& 0.052 &  --   & 0.073 &  --   & 0.108 & 0.715 & 0.152 & 0.185 & 0.75   &  1.10 &  1.06   &   1.19       &  --  \\[0.1cm]    
                               &  m  &$\times$& 0.154 &  --   & 0.091 &  --   & 0.118 & 0.372 & 0.170 & 0.192 & --     &  --   &  --     &   --         &  --  \\[0.1cm] 
\hline
\multirow{2}{*}{+(0.175--0.275)} &  l  &$\times$& 0.050 &  --   & 0.111 &  --   &  --   & 0.745 & --    & 0.092 & 0.63   & 0.57  &  0.53   &   1.05       &  --  \\[0.1cm]   
                               &  m  &$\times$& 0.057 &  --   & 0.115 &  --   &  --   & 0.656 & --    & 0.126 & --     & --    &  --     &   --         &  --  \\[0.1cm]  
\hline 
\multirow{2}{*}{+(0.075--0.175)} &  l  & 0.136  & --    &  --   & --    & 0.028 & 0.054 & 0.668 & 0.023 & 0.043 &  0.55  & 0.53  &  0.55   &   1.04       & 0.57 \\[0.1cm]    
                               &  m  & --     & --    &  --   & --    & 0.045 & 0.143 & 0.660 & 0.062 & 0.095 &  --    & --    &  --     &   --         & --   \\[0.1cm]  
\hline 
\multirow{2}{*}{+(0.025--0.075)} &  l  & 0.140  & --    &  --   & --    & 0.032 & 0.070 & 0.659 & 0.101 & 0.005 &  0.51  & 0.50  &  0.53   &   1.03       & 0.54 \\ [0.1cm]   
                               &  m  & --     & --    &  --   & --    & 0.078 & 0.098 & 0.563 & 0.157 & 0.014 &  --    & --    &  --     &   --         & --   \\ [0.1cm] 
\hline
\multirow{2}{*}{--0.025 -- +0.025}&  l & 0.170  & --    &  --   & --    & 0.026 & 0.029 & 0.634 & 0.152 & --    &  0.41  & 0.36  & 0.40    &   0.98       & 0.39 \\ [0.1cm]   
                               &  m  & --     & --    &  --   & --    & 0.028 & 0.034 & 0.735 & 0.203 & --    &  --    & --    & --      &   --         & --   \\ [0.1cm] 
\hline
\multirow{2}{*}{--(0.025--0.075)}&  l  & 0.156  & --    &  --   & --    & 0.034 & 0.027 & 0.648 & 0.106 & 0.019 &  0.45  &  0.40 & 0.44    &   1.02       & 0.43 \\ [0.1cm]   
                               &  m  & --     & --    &  --   & --    & 0.030 & 0.025 & 0.738 & 0.202 & 0.005 &  --    &  --   & --      &   --         & --   \\ [0.1cm] 
\hline
\multirow{2}{*}{--(0.075--0.175)}&  l  & 0.142  & --    &  --   & --    & 0.057 & 0.063 & 0.599 & 0.077 & 0.052 &  0.48  & 0.49  & 0.47    &   1.03       & 0.52 \\ [0.1cm]   
                               &  m  & --     & --    &  --   & --    & 0.068 & 0.094 & 0.686 & 0.125 & 0.083 &  --    & --    & --      &   --         & --   \\ [0.1cm] 
\hline 
\multirow{2}{*}{--(0.175--0.275)}&  l  &$\times$& 0.035 &  --   & 0.128 & --    & 0.072 & 0.691 & --    & 0.074 &  0.51  &  0.50 &   0.48  &   1.05       &  --  \\ [0.1cm] 
                               &  m  &$\times$& 0.021 &  --   & 0.105 & --    & 0.093 & 0.619 & --    & 0.162 &  --    &  --   &   --    &   --         &  --  \\ [0.1cm] 
\hline
\multirow{2}{*}{--(0.275--0.375)}&  l  &$\times$& 0.109 & --    & 0.093 & --    & 0.145 & 0.481 & --    & 0.173 &  0.59  &  0.56 &  0.58   &   1.12       &  --  \\ [0.1cm]   
                               &  m  &$\times$& 0.005 & --    & 0.174 & --    & 0.297 & 0.386 & --    & 0.368 &  --    &  --   &  --     &   --         &  --  \\ [0.1cm] 
\hline
\multirow{2}{*}{--(0.375--0.475)}&  l  &$\times$& 0.155 & 0.187 & 0.184 & --    & 0.084 & 0.250 & --    & 0.279 &  0.71  &  0.63 &  0.65   &   1.17       &  --  \\ [0.1cm]   
                               &  m  &$\times$& 0.094 & 0.092 & 0.073 & --    & 0.148 & 0.261 & --    & 0.332 &  --    &  --   &  --     &   --         &  --  \\ [0.1cm] 
\hline
\multirow{2}{*}{--(0.475--0.575)}&  l  &$\times$& 0.068 &  --   & 0.180 & 0.109 & 0.264 & 0.220 & 0.160 & --    &  0.88  &  0.45 &   0.47  &   1.25       &  --  \\[0.1cm] 
                               &  m  &$\times$& 0.009 &  --   & 0.171 & 0.076 & 0.267 & 0.375 & 0.102 & --    & --     & --    &   --    &   --         &  --  \\[0.1cm] 
\hline   
\multirow{2}{*}{--(0.575--0.675)}&  l  &$\times$& --    &  --   & 0.129 & 0.309 & 0.310 & 0.073 & --    & 0.179 &  1.03  &  0.47 &   0.50  &   1.31       &  --  \\[0.1cm] 
                               &  m  &$\times$& --    &  --   & 0.010 & 0.315 & 0.386 & 0.129 & --    & 0.259 & --     &  --   &   --    &   --         &  --  \\[0.1cm] 
\hline   
\multirow{2}{*}{--(0.675--0.825)}&  l  &$\times$& --    & 0.084 & 0.066 & 0.331 & 0.316 & 0.160 & --    & 0.042 &  0.98  &  0.38 &   0.36  &   1.43       &  --  \\[0.1cm] 
                               &  m  &$\times$& --    & 0.110 & 0.021 & 0.163 & 0.250 & 0.161 & --    & 0.286 & --     &  --   &   --    &   --         &  --  \\[0.1cm]
\hline \hline 
\end{tabular}
\tablenotemark{}
\tablecomments{Best-fit stellar population models of binned STIS spectrum.  Column 1: radii included in bin; particularly, the five central most data points are single pixels, while the bins at larger radii combine multiple pixels. We therefore use the notation of $\pm$(r$_1$--r$_2$)$\arcsec$ to specify the size of those spectra bins. Column 2: l (light fraction) and m (mass fraction) for each best-fit model. Column 3: AGN component; $\times$ means no AGN component was included in the fit. Column 3 -- 12: SSP model age light and mass fractions. Column 13: Attenuation of starlight by dust. Column 14 -- 15: \ml$_I$ for each spectral fit using the BC03 and CB07 SSP models. Column 16: reduced $\chi^2$ of each bin associated with BC03 SSP model. Column 17: the best-fit \ml$^{\star}$~assuming no AGN contribution for the five central spectral bins.}
\label{tab_bin_ssp}
\end{table*}
\begin{figure*}[!htb]
\minipage{0.5\textwidth}
  \includegraphics[width=\linewidth]{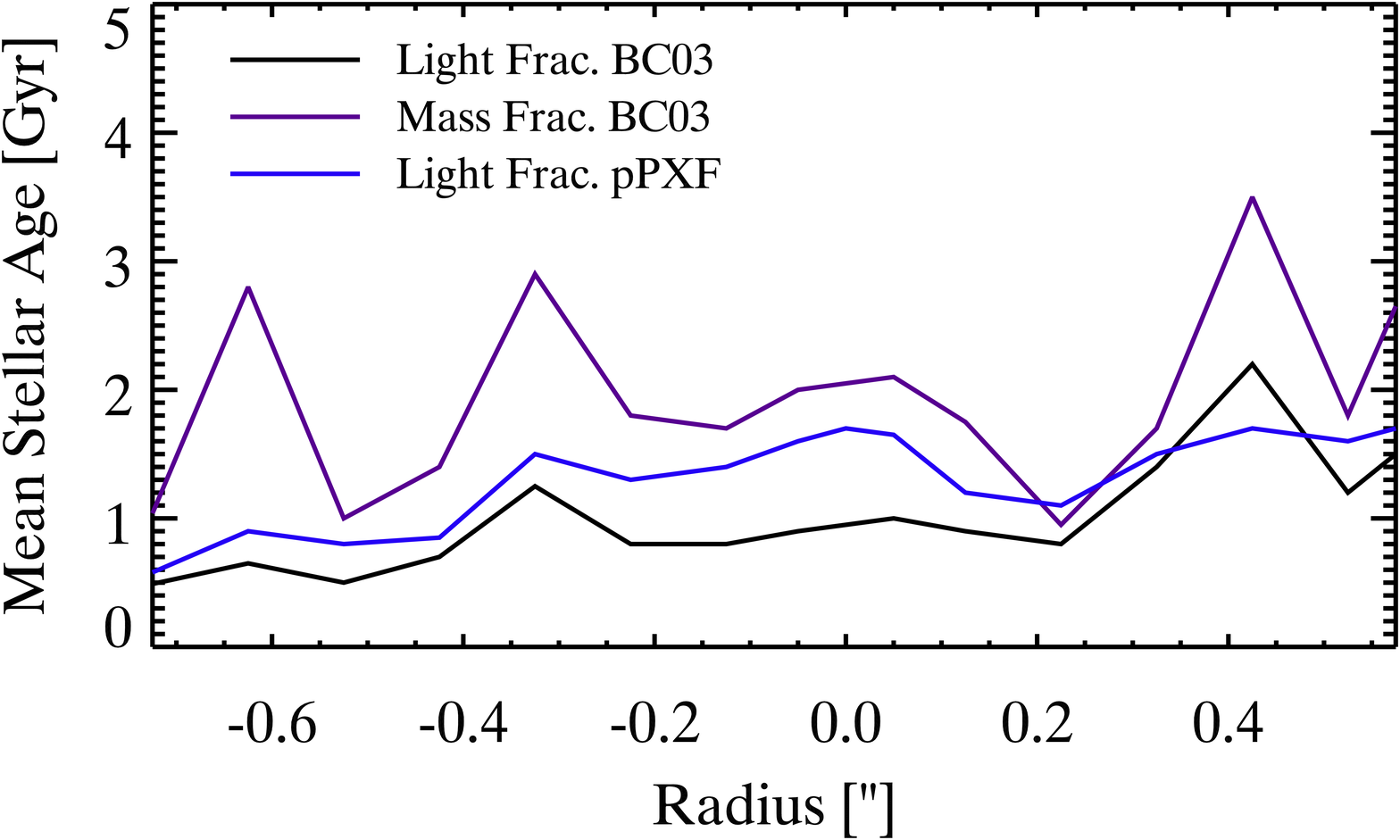}\label{mean_stellar_age}
\endminipage\hfill
\minipage{0.5\textwidth}
  \includegraphics[width=\linewidth]{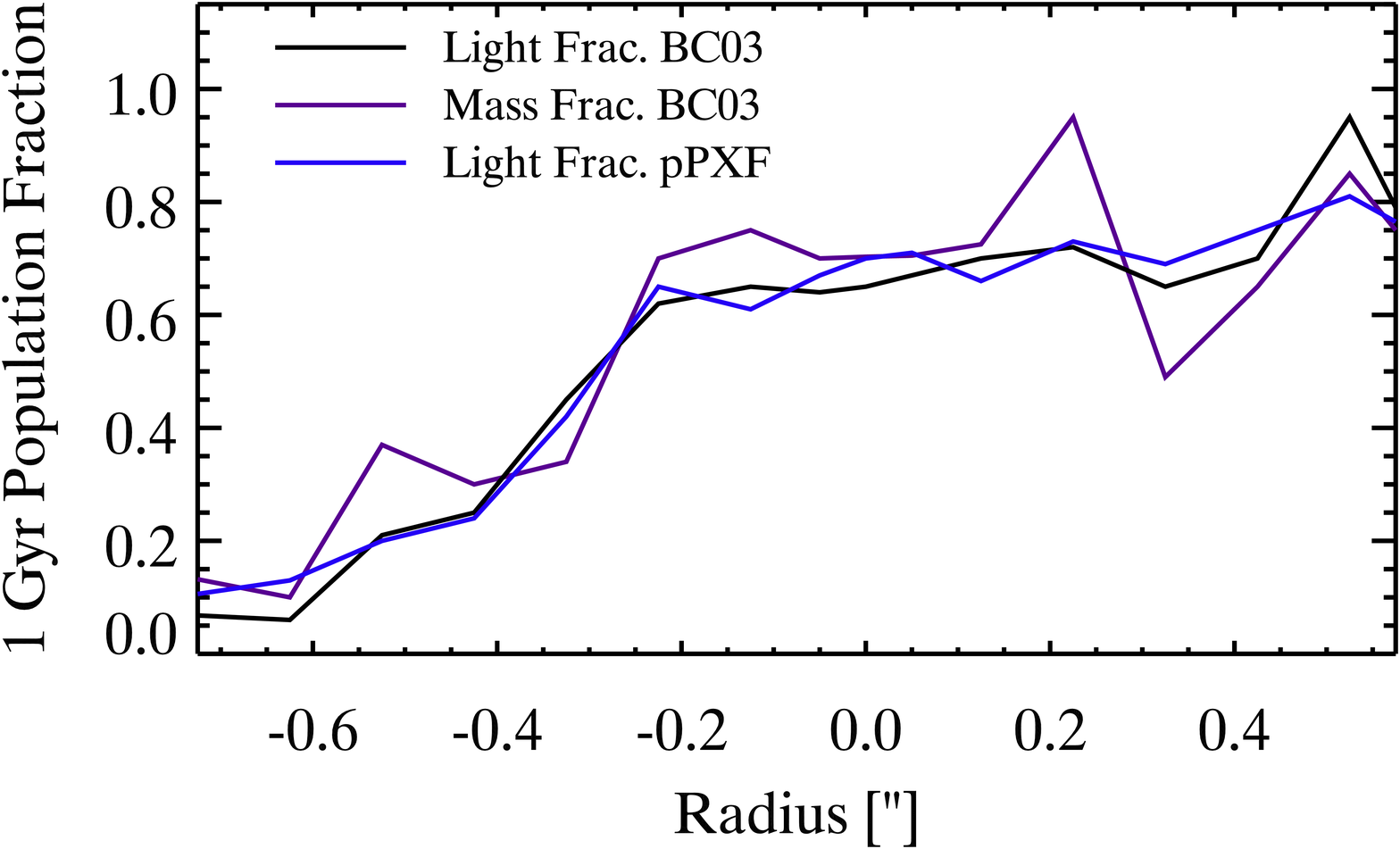}\label{1Gyr_pop_frac}
\endminipage
  \caption[SCP06C1]{{\em Left panel:} the light- and mass-weighted mean stellar age of the NSC vs. radius as derived from model fits to the STIS spectra.  {\em Right panel:} the light- and mass-fractions of the 1~Gyr population in each of these fits; this population dominates the nuclear region, but contributes less at larger radii (S10). The pPXF light fraction results are also presented to demonstrate the level of consistency with our BC03 models.}
\label{1Gyr_pop}
\end{figure*}

Based on the evidence for variability we found in the previous section, it appears an accreting BH may contribute to the continuum flux near the center of the galaxy.  We therefore test whether including a power-law AGN component provides a better fit to the STIS data of the nucleus.  We fit the nuclear spectrum ($\leq 0\farcs05$) using the SSPs described both with and without an AGN continuum component: $f_{\lambda}\sim\lambda^{-\alpha}$. To determine the best fit value of $\alpha$, we fit the spectrum over the entire STIS wavelength range; $\alpha$ is varied from -4.0 to 4.0 in steps of 0.1, and the resulting chi-square is evaluated for each model. We find the best fit of $\alpha=0.5^{+0.4}_{-0.3}$; as shown in Figure~\ref{agn_ssp}, these values provide the best fit over the full wavelength range, and also in the UV.  The left panel of Figure~\ref{agn_ssp} shows the $\chi^2$ of the whole spectrum as well as the $\chi^2$ in just the UV part of the spectrum  ($\lambda<4,000$~\AA) where the AGN is expected to contribute the largest fraction of the flux.

We then compare a fit with an AGN component with $\alpha=0.5$ to one without an AGN component and find that the inclusion of an AGN component provides a better fit.  The residuals of these two fits are shown in the right panel of Figure~\ref{agn_ssp}; while both provide visually similar fits, some clear differences are seen. Most notably, the Balmer emission lines in the non-AGN case are stronger and show a similar strength at higher orders (an unphysical scenario). More quantitatively, the non-AGN Balmer line ratios are H$\gamma$/H$\beta\sim0.48$, H$\delta$/H$\beta\sim0.34$, and H$\epsilon$/H$\beta\sim0.49$. In the AGN case, the higher order Balmer lines are weaker with H$\gamma$/H$\beta\sim0.47$, H$\delta$/H$\beta\sim0.24$, and H$\epsilon$/H$\beta\sim0.15$.  These ratios are consistent with \citet{Osterbrock89} (Case A recombineation with T$\sim$5,000 K) which gives H$\gamma$/H$\beta$~$\sim$~0.458, H$\delta$/H$\beta$~$\sim$~0.250, and H$\epsilon$/H$\beta$~$\sim$~0.153. We note however that similar line-ratios to the non-AGN fits are seen at larger radii where a power law component would not be expected (e.g.,~Figure~\ref{ssp_fit}, right panel).

We note that these emission lines were excluded from the fit and thus do not contribute to the $\chi^2$ value, but the left panels show the improved $\chi^2$ and residuals for the fit including an AGN component.  To quantify how significant this difference is, we calculate the Akaike Information Criterion \citep{Cavanaugh97, Burnham02} corrected for finite sample sizes (AICc). Given the best-fit $\chi^2$ value of 206 for the model with an AGN and $\chi^2$ of 209 for the model without an AGN, the AICc suggests that the AGN model is $\sim$5 times more probable than the no AGN model.  While this alone is not compelling evidence for an AGN component, the somewhat better fits combined with the strong evidence for nuclear variability presented in the previous subsection lead us to include an AGN power-law component for our population synthesis fits at radii \textless0$\farcs$2.  

\begin{figure*}[!htb]
\minipage{0.5\textwidth}
  \includegraphics[width=\linewidth]{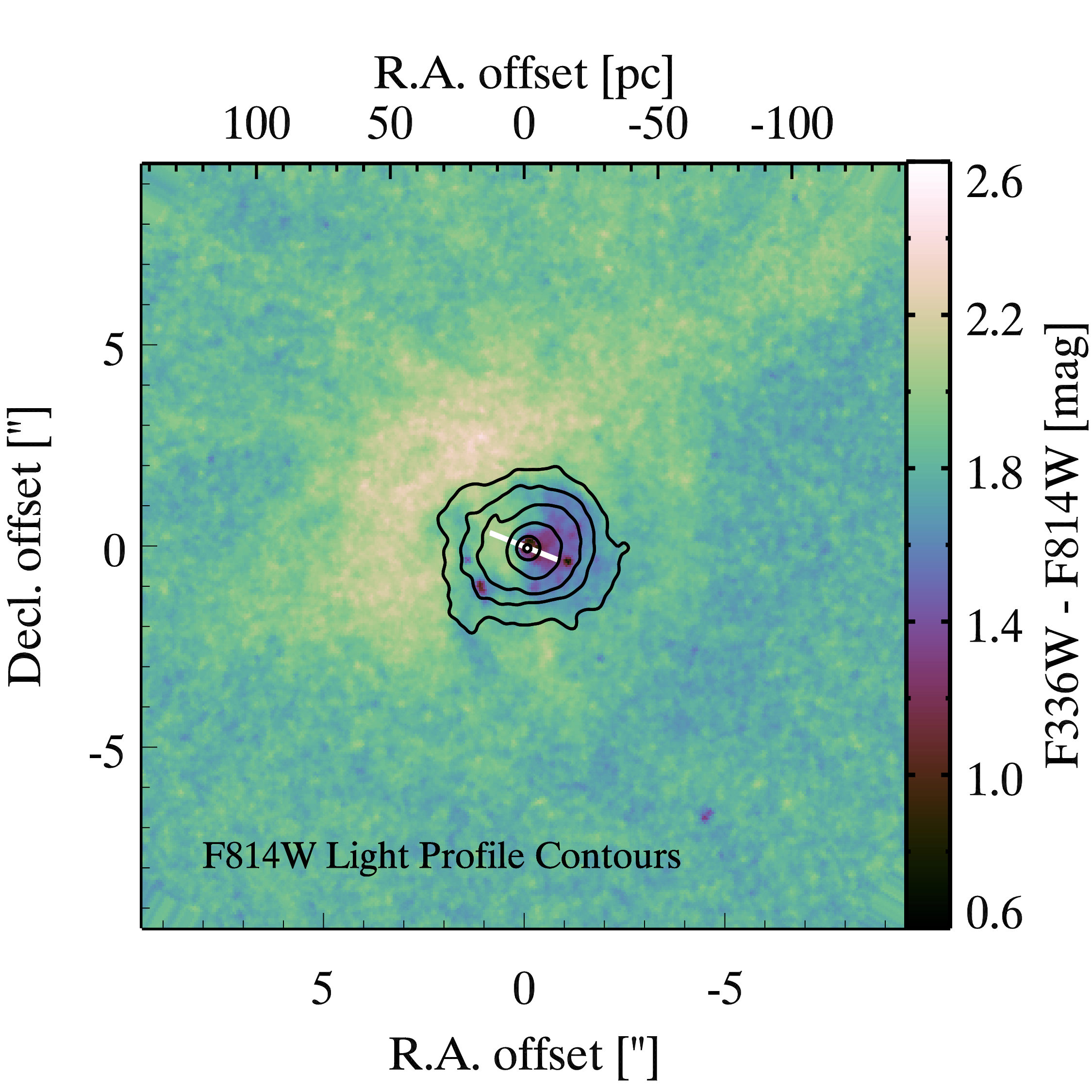}\label{colormap1}
\endminipage\hfill
\minipage{0.5\textwidth}
  \includegraphics[width=\linewidth]{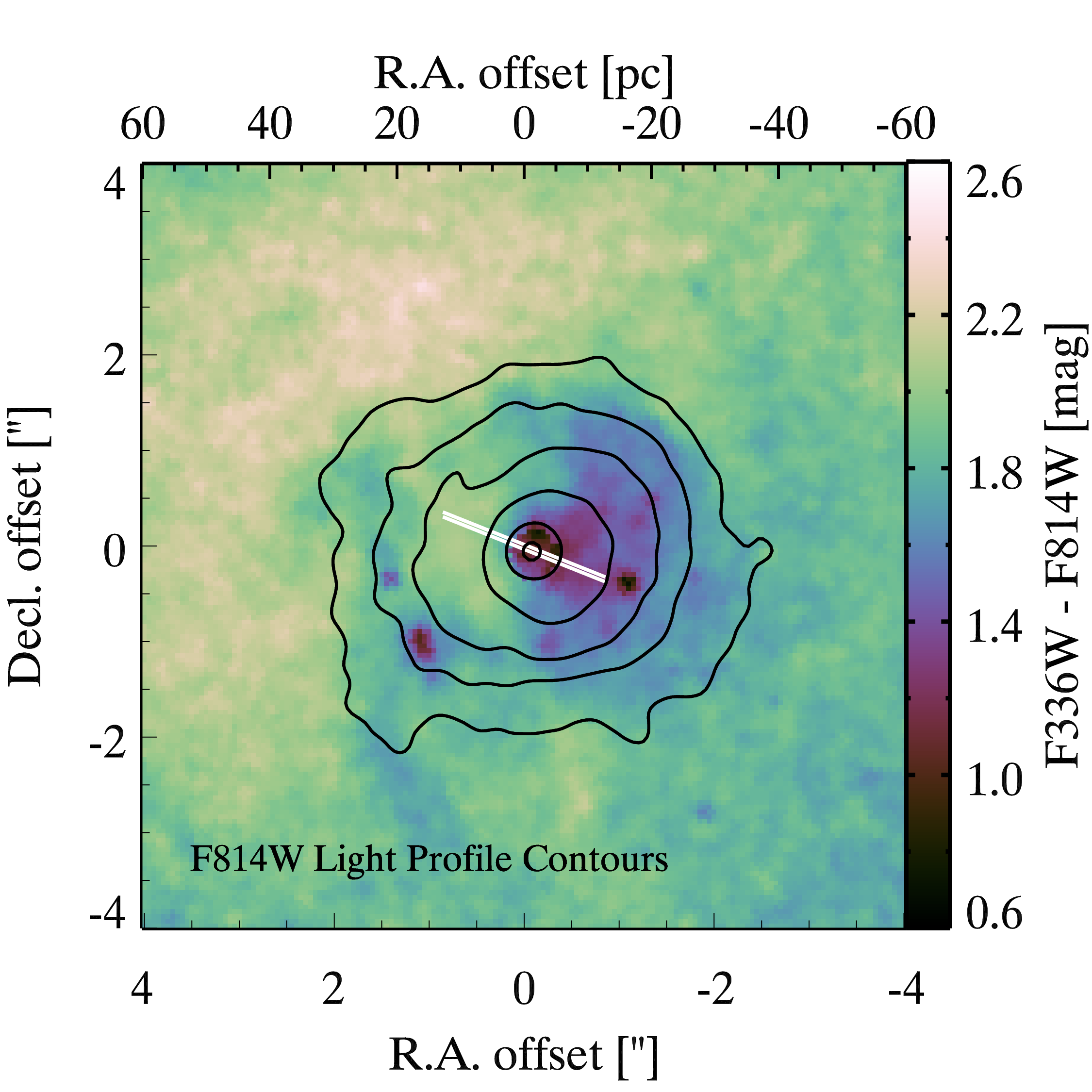}\label{colormap2}
\endminipage
\caption{ Color maps of the nucleus at two different resolutions.  The color map shown is the \FB--\FI~map that has been cross-convolved to match PSFs. The contours show the \FI~surface brightness at $\mu_{F814W}$ of 13.2, 13.6, 14.0, 14.5, 15.1 mag arcsec$^{-2}$.  The location of the STIS spectrum is shown with white parallel lines and extends out to the radius at which stellar population information is available.  The redder regions result from dust extinction, while the western half of the nucleus appears to have little internal extinction. The center of NGC~404 nuleus is represented as (0, 0)$\arcsec$ corresponding to R.A. = 01$^{\rm h}$09$^{\rm m}$27$^{\rm s}.01$ and Decl. = 35\deg43$^{\prime}$4$^{\prime\prime}$.8.}
\label{colormap}
\end{figure*}

\begin{figure}[ht]
     \centering
      \epsscale{1.2}
          \plotone{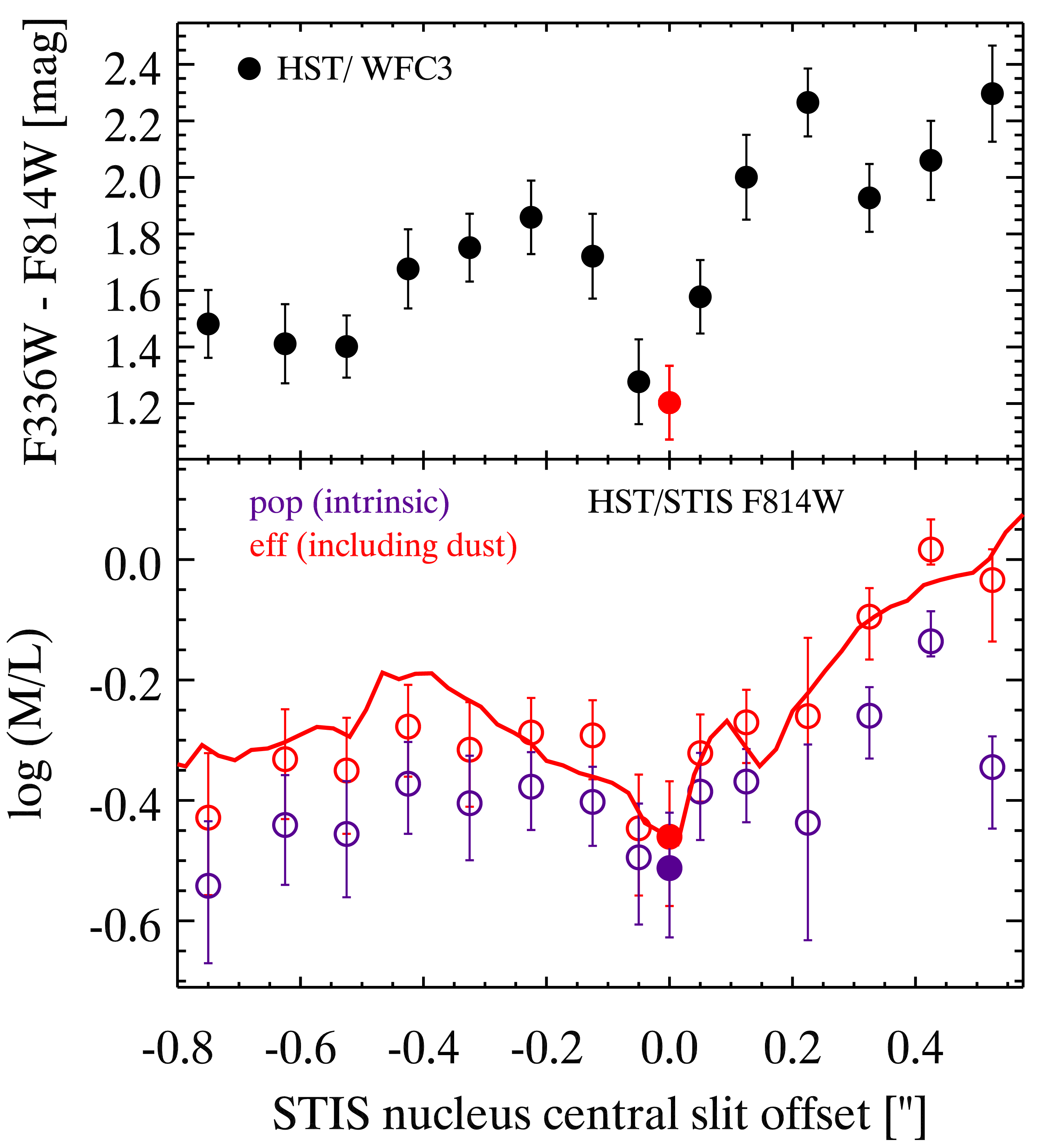}
          \caption[SCP06C1]{ The color and~\ml~along the STIS slit. {\em Top panel:} the variation in \FB-\FI~color along the STIS slit. The central point is shown with a red dot.  The color was determined by astrometrically aligning the WFC3 images to the STIS spectroscopy. {\em Bottom panel:}  The~\ml~values in the \FI~band derived from the stellar population fits to the STIS data.  Both the~\mlpop~(purple; not including dust extinction) and the~\mleff~(red; including dust extinction) are shown.  The filled circles indicate the central pixel.  The red line shows the~\mleff~reconstructed from the~\mleff~vs. color correlation (as shown in Figure~\ref{color_m2leff_correlation}).} 
\label{nucleus_color_m2l_vary}
\end{figure}

{\em Nuclear Line Emission:} From the residual spectrum of the central three pixels ($\leq$$0\farcs05$ or 0.75~pc) of the STIS long-slit spectroscopy we fit the Balmer H${\beta}$ recombination emission line with a Gaussian and find a flux of $(2.8\pm0.3)\times10^{-16}$ erg s$^{-1}$ cm$^{-2}$.  In S10, the total nuclear H$\beta$ flux in an aperture of $1\farcs0\times2\farcs3$ was $\sim1.08\times10^{-13}$ erg s$^{-1}$ cm$^{-2}$, thus the fraction of H$\beta$ flux in the nuclear spectrum is only 0.2\% of the total emission in the nuclear region.

From the nuclear H${\beta}$ flux we can estimate the nuclear H${\alpha}$ flux to use for comparison to the X-ray luminosity and to estimate a nuclear star formation rate.  We assume the Balmer decrement measured by \citet{Osterbrock89} of H$\alpha$/H$\beta$ $\sim 3.1$ for the case of AGN, giving an H$\alpha$ flux of $(8.72\pm0.05)\times10^{-16}$ erg s$^{-1}$ cm$^{-2}$ and $L_{\rm H\alpha}=(9.9\pm0.1)\times10^{35}$ erg s$^{-1}$ with \citet{Cardelli89} interstellar extinction law. In the case of star formation, H${\alpha}$/H$\beta$ $\sim 2.85$, giving an H$\alpha$ flux of $(7.9\pm0.2)\times10^{-16}$ erg s$^{-1}$ cm$^{-2}$ and $L_{\rm H\alpha}=(7.8\pm0.1)\times10^{35}$ erg s$^{-1}$.

\citet{Binder11} detected an X-ray point source associated with the NSC with $L_{\rm 2 - 10 keV}=1.3^{+0.8}_{-0.5}\times10^{37}$ erg s$^{-1}$.  We compare the nuclear $L_{\rm H{\alpha}}$ to this X-ray luminosity to determine its consistency with AGN emission.  From this H$\alpha$ flux, we can predict the X-ray flux using the empirical relationship of \citet{Panessa06}, with revised fit by \citet{Nguyen14}.  The observed  X-ray luminosity is consistent with the predicted flux from this relationship ($L_{\rm 2 - 10 keV}=8.7^{+58.4}_{-5.0}\times10^{36}$ erg s$^{-1}$).  The uncertainty in the prediction here is very large but shows that the emission line signal we see is consistent with an AGN source to the X-ray emission.

\begin{figure}[h]
     \centering
      \epsscale{1.2}
          \plotone{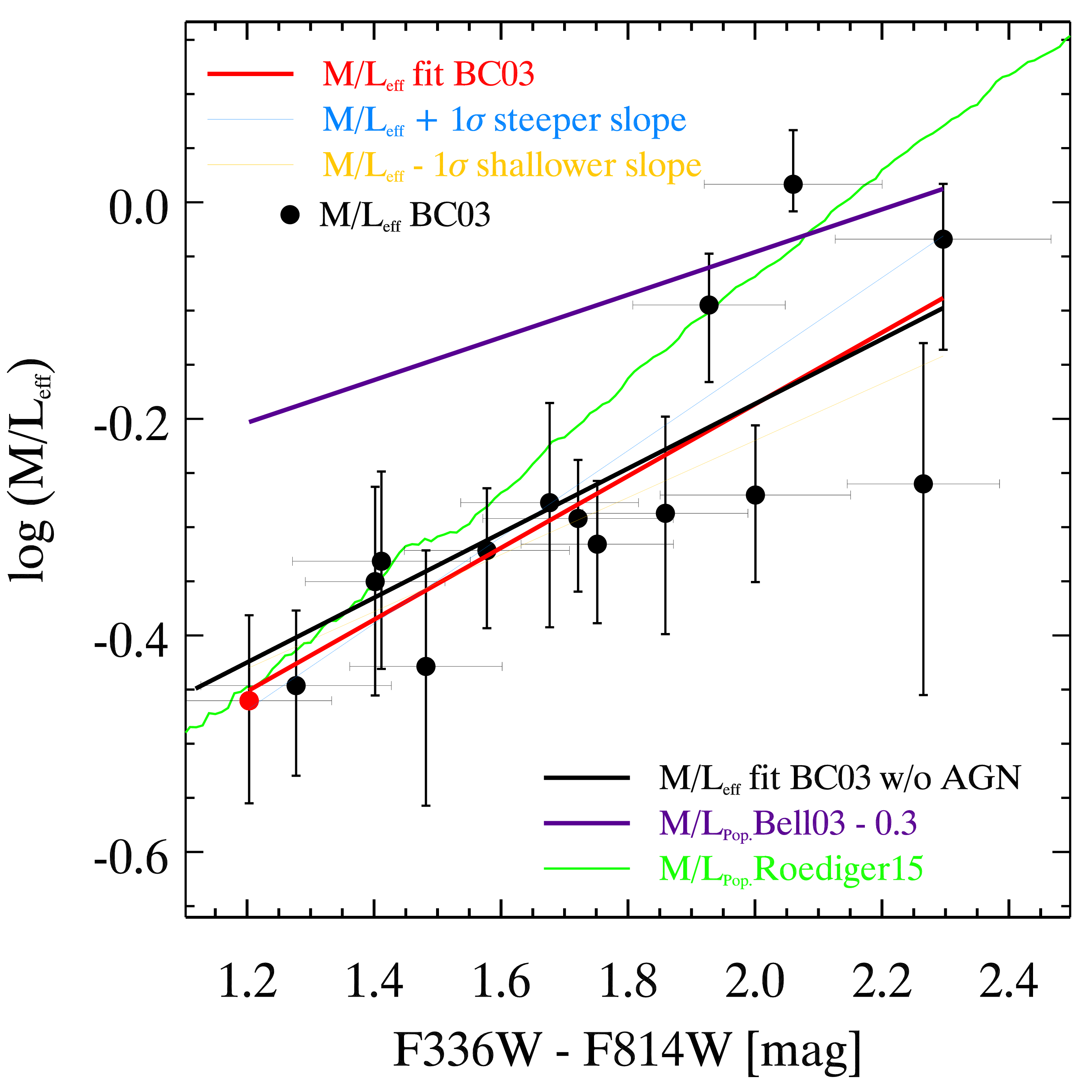}
          \caption{The mass-to-light ratio -- color relation for the NGC~404 nucleus.  The y-axis shows the effective mass-to-light ratio in \FI~(\mleff) determined from stellar population fits to the STIS spectroscopy, while the x-axis shows the \FB--\FI~color determined from WFC3 imaging.  Black points show the data for the stellar population fits using the BC03 models, while the thick red line shows the best-fit linear relation to these data. The 1$\sigma$ uncertainties in the slope of our best-fit linear relation are shown with thin blue (steeper) and yellow (shallower) lines.  The black thin line shows the best-fit linear relation to the data when we don't include an AGN component in the five innermost spectral bins during the fitting procedure. Error bars for each point in log(\mleff) were determined through a Monte Carlo analysis of the stellar population fits and these errors form the dominant error in our best-fit log(\mleff) color relations. The purple thick solid line is the predicted color--\ml~correlation from \citet{Bell03} shifted downwards by 0.3 dex, while the green line shows the \citet{Roediger15} relation.  The red data point indicates the NGC 404 center.}
\label{color_m2leff_correlation}
\end{figure}   

If we instead assume the emission is coming from star formation, we can use the H${\alpha}$ luminosity $L_{\rm H{\alpha}}=(7.8\pm0.1)\times10^{35}$ erg s$^{-1}$ to estimate an upper limit on the nuclear star formation rate.  Using the relation of \citet{Murphy11}, we find a nuclear SFR of $(4.2\pm0.4)\times10^{-6}$\Msun~yr$^{-1}$.  This SFR translates into an expected radio luminosity at 5~GHz of $(1.6\pm0.2)\times10^{22}$ erg s$^{-1}$~Hz$^{-1}$ \citep{Murphy11}; this is significantly lower than the observed total radio luminosity of $3.4\times10^{24}$ erg s$^{-1}$~Hz$^{-1}$ at the center of NGC~404 (Nyland et al. 2016, {\em in prep}).  The radio morphology shows a compact source consistent within astrometric errors of the photometric and kinematic center of our NIFS data; the unresolved fraction of the total flux is $\sim$75\%, thus, this emission is still far too large to be related to the H$\alpha$ emission via star formation.  Thus, the nuclear H$\alpha$ emission favors AGN-related nuclear X-ray and radio emission.

\subsubsection{Stellar Population Modeling of STIS Data}\label{sssec:stellar_pop}

To obtain sufficient S/N (\textgreater20) for stellar population modeling, we bin the STIS spectrum along the slit.  Near the center, the individual pixels ($0\farcs05$) exceed this S/N threshold, but the outermost bins (at $r=0\farcs675-0\farcs825$) extend over $0\farcs15$.  All of the bins analyzed here had S/N~$> 20$ at wavelengths longer than 5,000\AA.  Bins with lower S/N appeared to have significant detector artifacts making stellar population fits unreliable. In this subsection, we focus on our best-fit results -- these include an AGN component in the central $0\farcs2$ and use the BC03 models.  We discuss the uncertainties in our modeling in the next section.

The best-fit model and residuals to the central pixel are illustrated in the left panel of Figure~\ref{ssp_fit}.   The model is shown in red with the individual SSP components shown in blue.  The details of these stellar populations are shown in Tables~\ref{tab_pop} and \ref{tab_bin_ssp}.   The fit in the central pixel is dominated by intermediate age stars with 70\% of the light in the 1 Gyr SSP, 5\% of the light in the 2.5 and 5~Gyr SSP, 17\% of the light in AGN (see Table~\ref{tab_pop}). Only a small fraction of the light is in younger populations (\textless570 Myr). The best-fit $A_V$ is 0.45 ($A_V=1.086\times\tau_V$) and the~\ml~values are 0.55, 0.36, and 0.29 in the $V$, $I$, and $H$ bands, respectively.  The right panel of this figure shows a lower S/N spectral bin spanning $+(0\farcs275-0\farcs375)$ with higher extinction; the best fit $A_V \sim0.93$ and the~\ml~values are 0.67, 0.52, and 0.38 in the $V$, $I$, and $H$ bands, respectively. The reduced $\chi^2$ is 0.98 for the central pixel and 2.23 for the outer bin; the decrease in the goodness-of-fit at larger radii is likely due to increased contributions from bad pixels. The details of SSP model fits to all the spectral bins are presented in Table~\ref{tab_bin_ssp}.

The left panel of Figure~\ref{1Gyr_pop} shows the light-weighted and mass-weighted ages along the slit. The mean light-weighted age is $\sim$2.5 Gyr and the mean mass-weighted age is $\sim$4 Gyr.  A large fraction of the light appears to be coming from a 1~Gyr population across the slit.  At radii $\lesssim$0$\farcs$3 ($\lesssim$6~pc) this stellar population contributes $\sim$70\% of the light and mass; this drops somewhat at larger radii (the right panel of Figure~\ref{1Gyr_pop}), continuing the trend seen at larger radii \citep{Bouchard10}.  This mass fraction is higher than the mass fraction of $\sim$1~Gyr found in the NSC by S10 using the same template fitting method using ground-based optical spectroscopy \citep[see also][]{CidFernandes04}, suggesting that this population is concentrated near the center of the NSC.  This central region corresponds to the ``extra-light'' counter-rotating component observed in S10 and provides additional evidence that this portion of the NSC was formed from externally accreted gas during a merger $\sim$1~Gyr ago.  This suggests that gas from minor mergers can effectively accrete into the very center of the galaxy to form an NSC. 

\begin{figure*}[!htb]
\minipage{0.5\textwidth}
  \includegraphics[width=\linewidth]{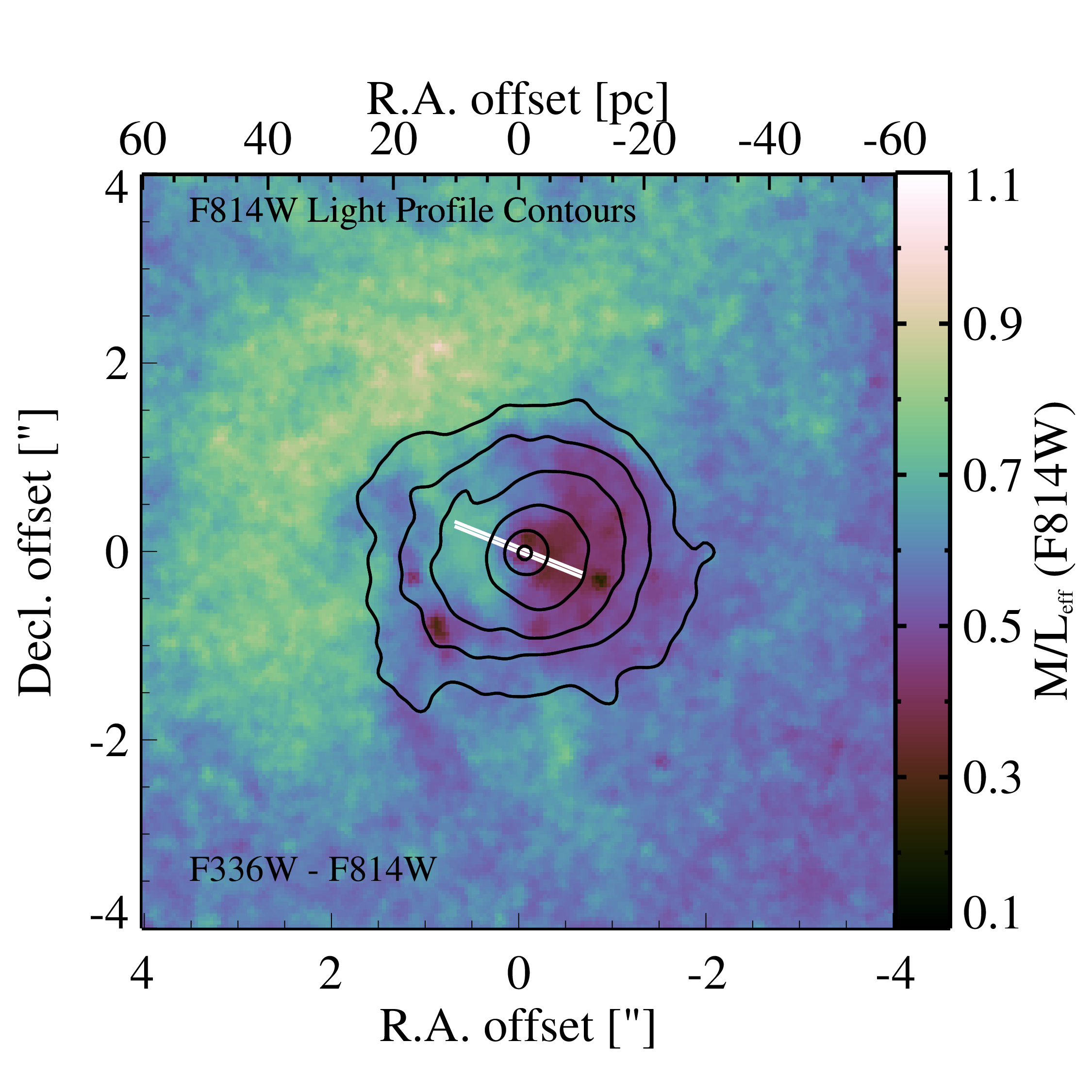}\label{m2l_map}
\endminipage\hfill
\minipage{0.5\textwidth}
  \includegraphics[width=\linewidth]{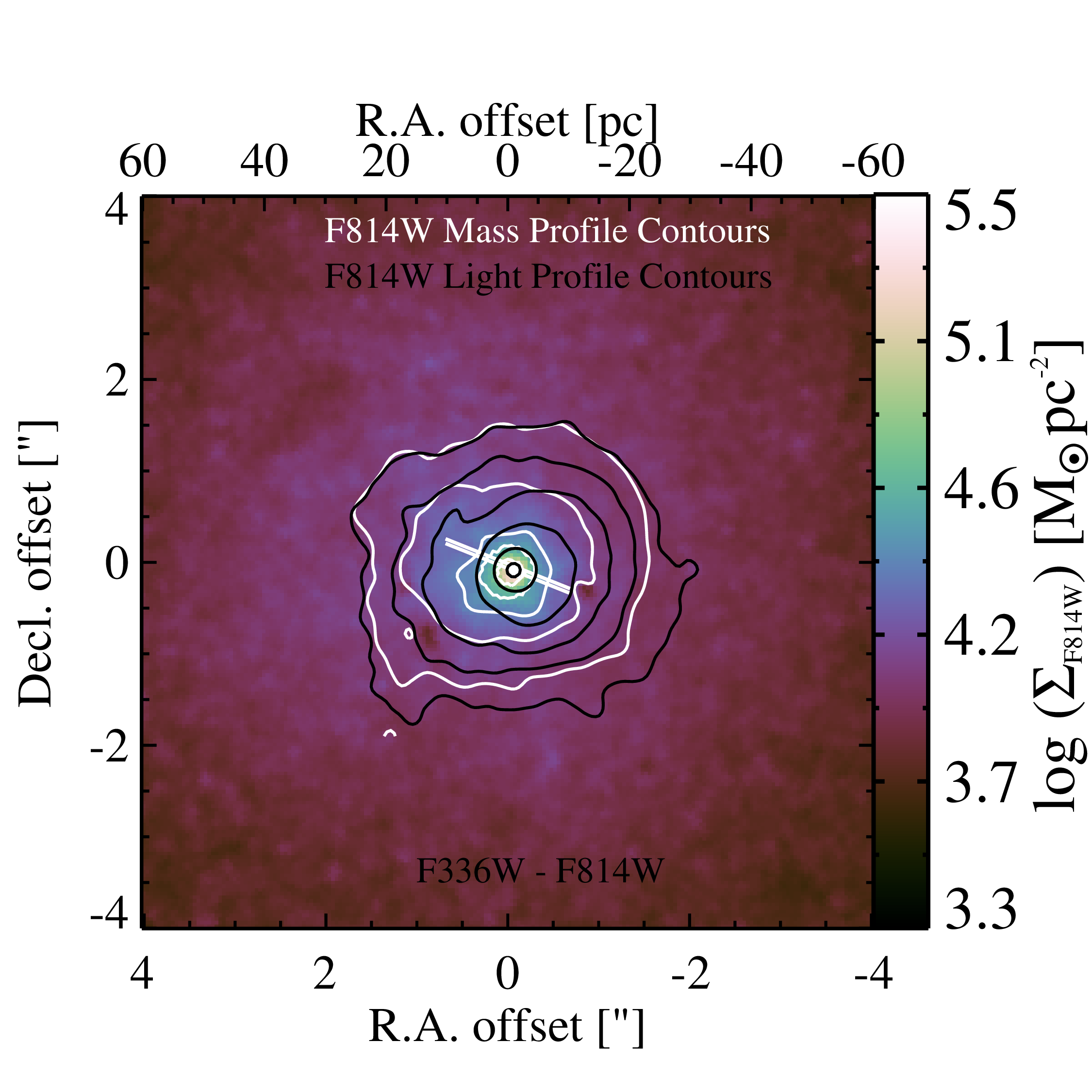}\label{mass_map}
\endminipage
\caption[SCP06C1]{{\em Left panel:}~\mleff~map of NGC~404 nucleus constructed using effective spectral~\mleff~vs. \FB--\FI~relation (Figure~\ref{color_m2leff_correlation}) to predict the \FI~\mleff. The black contours and slit are as in Figure~\ref{colormap}. {\em Right panel:} The mass surface density map of the nucleus created by multiplying the \FI~image with the the \mleff~map shown in the left panel.   The white contours show the mass profile at 5.5$\times10^5$, 5.1$\times10^5$, 6.5$\times10^4$, 4.0$\times10^4$, and 1.0$\times10^4$\Msun~pc$^{-2}$, while the black contours are the same as in Figure~\ref{colormap}.}
\label{m2l_mass_map}
\end{figure*}

\subsubsection{Assessing Uncertainties in the Stellar Population Modeling}\label{sssec:systemic_error} 

\begin{table*}[ht]
\caption{\textsc{galfit} Mass Map Model Profile in WFC3 \FI\tablenotemark{a}}
\centering
\begin{tabular}{cccccccc} 
\hline \hline 
Component    &  Label     & S\'ersic $n$& $r_{\text{eff}}$ & $r_{\text{eff}}$ &    P.A.       &     $b/a$       & $M$                \\
                       &               &                   &  [${\rm pc}$]     & [$\arcsec$]     & [$^{\circ}$] &                   &[$\times10^6$\Msun] \\
  (1)                 & (2)         &      (3)       &         (4)    &      (5)    &  (6)            &    (7)             &      (8)      \\
\hline     
Central excess S\'ersic  & Central S & 0.5$\pm$0.1 &  1.6$\pm$0.1   & 0.11$\pm$0.01  & 20.7$\pm$0.8& 0.974$\pm$0.006 & 3.4$\pm$0.2        \\[0.1cm] 
Inner S\'ersic           & Inner S      &1.96$\pm$0.17&  20.1$\pm$0.4  & 1.33$\pm$0.04  & 60.0$\pm$3.7& 0.958$\pm$0.004 & 10.1$\pm$0.1       \\[0.1cm]   
Outer S\'ersic (Bulge)  & Outer S   &     2.5     &      675       &       45       &     80      & 0.997$\pm$0.003 &844.2$\pm$6.4       \\[0.1cm]          
\hline 
\end{tabular}
\footnotetext[1]{Parameter values without error bars means that we fix those parameters during \textsc{galfit}.}
\tablenotemark{}
\tablecomments{\textsc{galfit} mass map model parameters of WFC3 \FI~mass map. Column 1 \& 2: component name and its label (see Figure~\ref{gal_mass_map_comps}). Column 3: component index profile. Column 4 and 5: effective radius or half-light (half-mass in the case of mass map) of the component profile in pc and $\arcsec$, respectively. Column 6: position angle. Column 7: ratio of major axis and minor axis. Column 8: mass of each component.}
\label{tab_sersic}  
\end{table*}

We estimate the uncertainty in the~\ml~of each spatial bin using a Monte Carlo technique.  Specifically, we add in Gaussian random errors to each spectrum and refit 100~times.  We then take the standard deviation of the resulting~\ml~distribution as the error on the~\ml~in that bin.   We find in the central bin, the~\ml~uncertainties are about 15\%, while at the edges the bins have~\ml~uncertainties of $\sim$30\%. We propagate these \ml~errors through the remainder of our analysis; these end up being the dominant error on the final mass maps we create.

We now discuss the sources of systematic uncertainties in our stellar population models.  The presence or absence of an AGN component significantly changes the youngest stellar populations, but has little effect on the derived \ml.  As shown in Table~\ref{tab_pop}, very young populations are inferred in the central pixel when no AGN is fit.  However, as discussed above, the featureless AGN spectrum appears to provide a better fit than these models.  We also note that the large 1~Gyr fraction is not strongly affected by our inclusion of AGN component.

We also find that BC03 and CB07 models provide very similar fits because the extra TP-AGB stars present in the CB07 models do not contribute significantly to the spectra short-wards of 6,000~\AA. The light- and mass-weighted ages are only different by 3--5\% between the CB07 and BC03 fits.  Because of this similarity, the~\ml~ratios in the $I$ band vary minimally, with maximum differences of $\sim$5\%.

To test our approach, we also fit our spectra with pPXF \citep{Cappellari04} using the MILES models for a subset of 50 logarithmically-spaced ages between 1~Gyr and 14~Gyr and seven metallicites from -2.32 to +0.22; this gives 350 model spectra \citep{Vazdekis10a, Vazdekis10b}.  We also include fits to gas emission lines. Because of a large number of templates, the resulting fit is highly non-unique; to handle this, we use regularization \citep[following][]{Onodera12,Cappellari13b,McDermid15}. Despite the wider range of templates, we find the pPXF star formation histories (SFHs) also feature a dominant intermediate age (1--3~Gyr) population presented as the blue line of Figure~\ref{1Gyr_pop}. These similarities also suggest our assumed mass-metallicity relationship has not dramatically affected our inferred SFHs.  Given the similarity of the fits, we proceed with our Simplefit results.

\subsection{Color Correlation With Spectroscopic M/L}\label{ssec:color_m2l} 

\begin{figure*}[!htb]
\minipage{0.33\textwidth}
  \includegraphics[width=\linewidth]{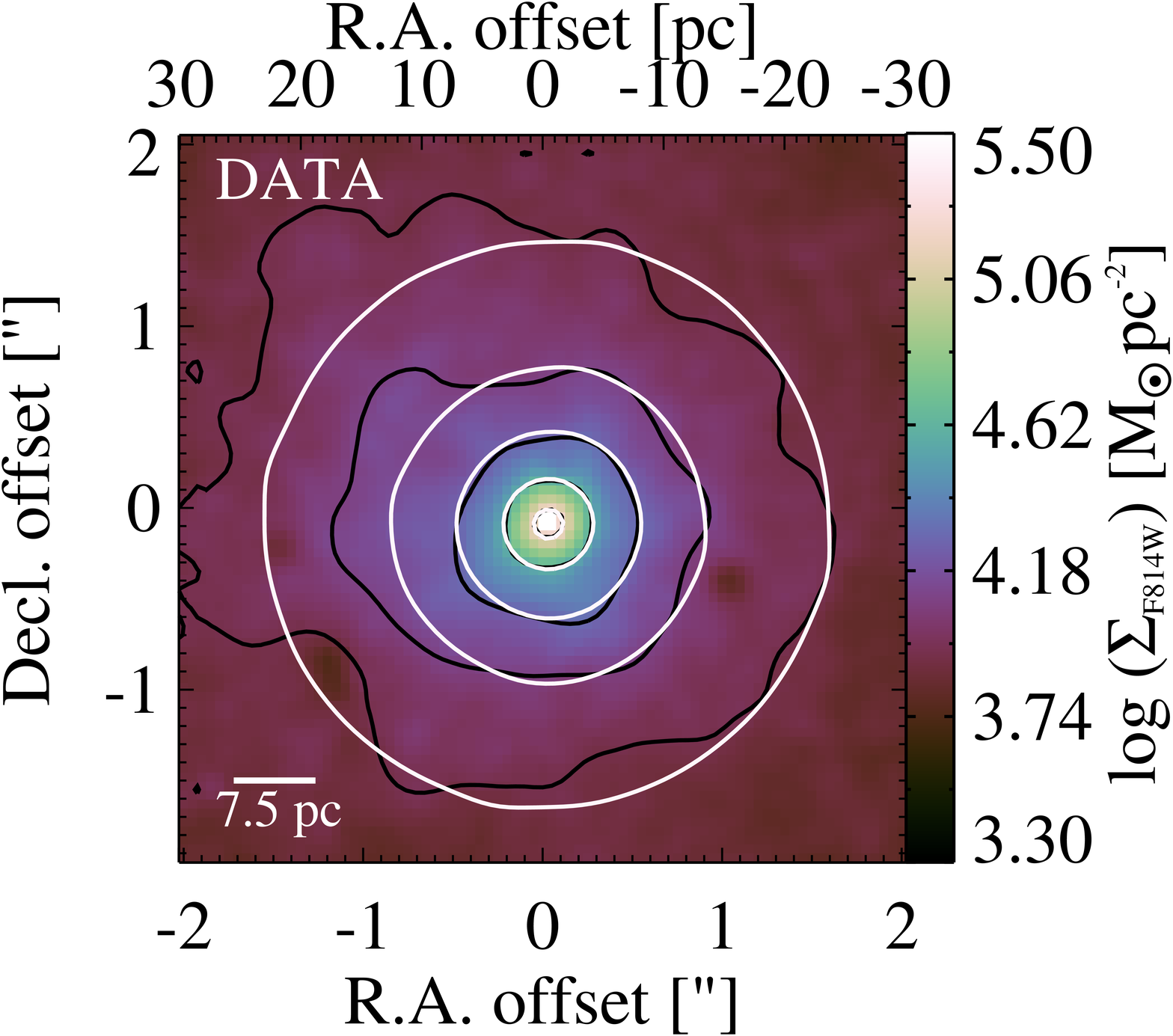}\label{hst_f814wmass}
\endminipage\hfill
\minipage{0.33\textwidth}
  \includegraphics[width=\linewidth]{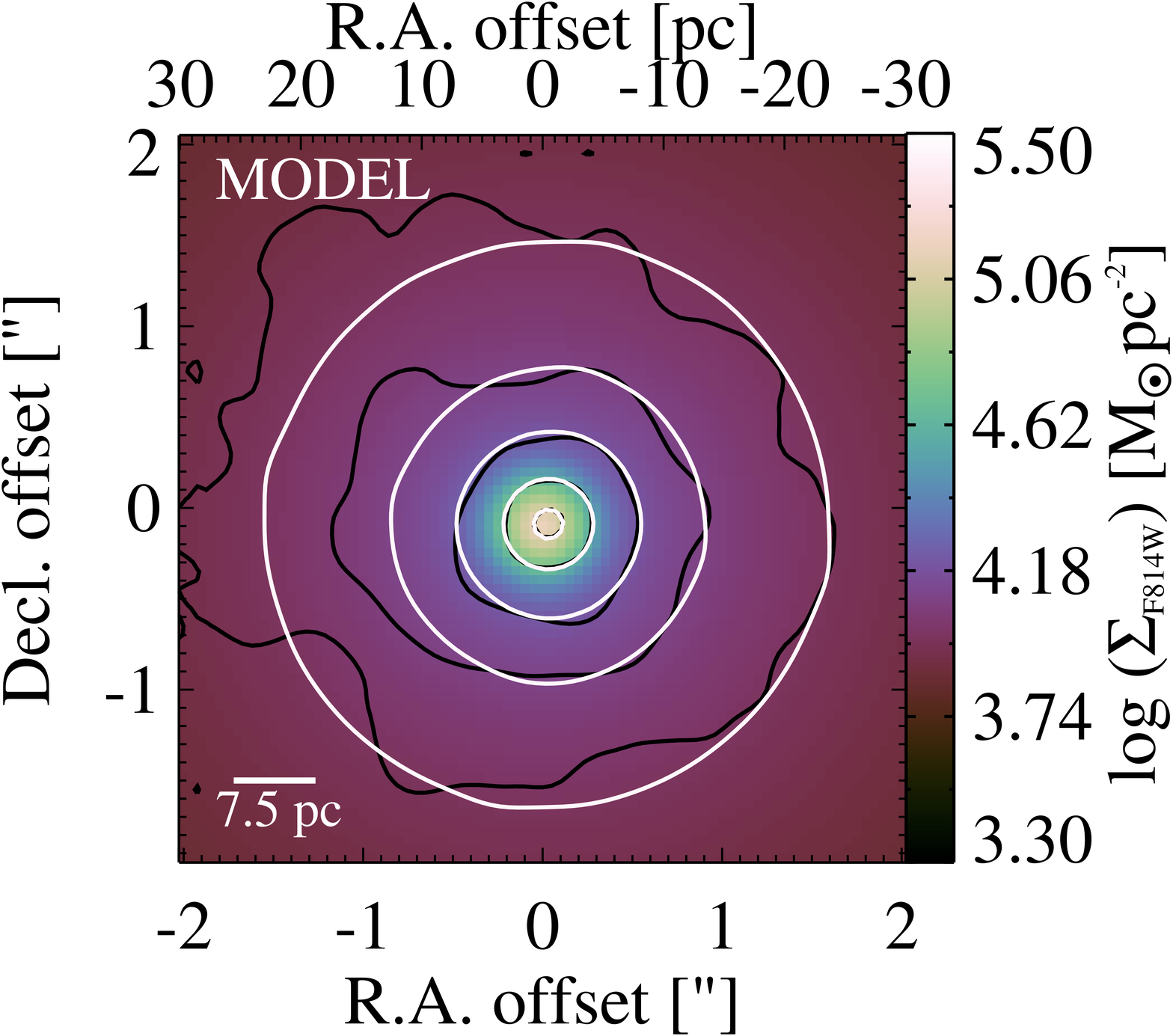}\label{gal_f814wmas}
\endminipage\hfill
\minipage{0.33\textwidth}
  \includegraphics[width=\linewidth]{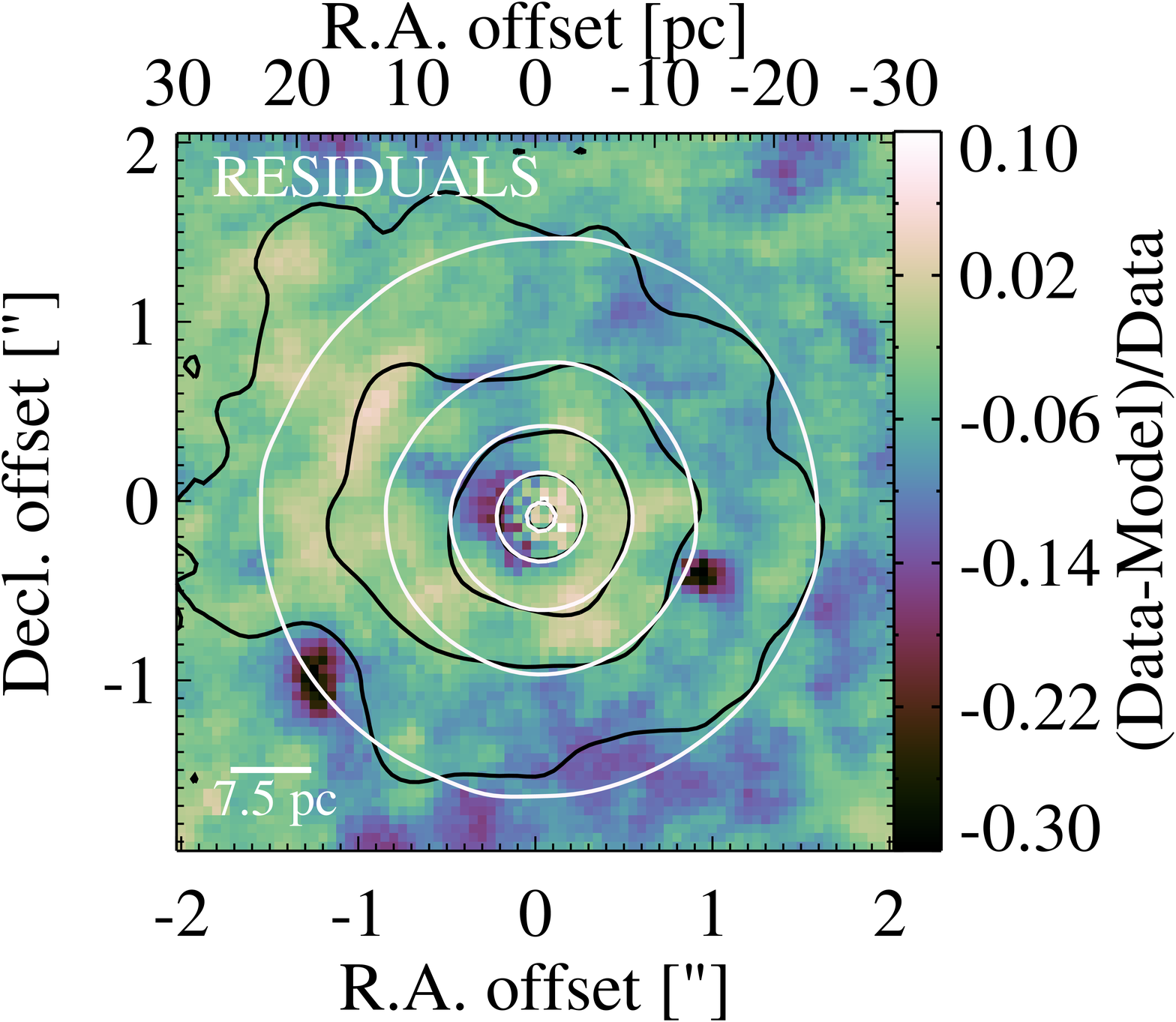}\label{res_f814wmass}
\label{res_hstgal_f814wmass}
\endminipage
\caption[SCP06C1]{\textsc{galfit} modelling of the mass surface density map.  {\em Left panel:} central portion of the mass surface density map shown in Figure~\ref{m2l_mass_map}.  {\em Middle panel:} the best-fit three-S\'ersic component \textsc{galfit} model to the the 2--D mass surface density map. {\em Right panel:} the fractional residual map ((Data-Model)/Data) of the mass surface density map as compared to the \textsc{galfit} model.  In each panel the black contours are created from the mass surface density map, while the white contours are created from the model map at the same levels as the black contours to highlight the consistency between the data and model.}
\label{galfit_mass_map} 
\end{figure*}

We examine the correlation between the WFC3 colors and STIS spectroscopic~\ml, and use this to create a new color--\ml~relation for the NGC~404 nucleus. This is a critical step in improving the BH mass estimate by more accurately modeling the mass distribution in the NGC~404 nucleus.

We create three WFC3 color maps of \FB--\FV, \FB--\FI, and \FV--\FI~by taking the astrometrically aligned image pairs and cross-convolving each image with the PSF of the other (e.g., for the \FB--\FV~color map, the \FB~image was convoluted with the \FV~PSF and vice versa).  The cross-convolution is necessary to accurately assess color gradients, since different width PSFs can introduce spurious gradients near the center of galaxies.  We then create color images using the Vega-based zero points and correcting for foreground extinction \citep[$A_{\rm F336W}=0.318$, $A_{\rm F547M}=0.194$, $A_{\rm F814W}=0.114$;][]{Schlegel98}. 
Figure~\ref{colormap} shows the WFC3 \FB--\FI~(approximately { \it U--I}) color map of the nucleus of NGC~404.   This map shows redder regions resulting from dust extinction on the eastern side of the nucleus while the western half of the nucleus appears to have little internal extinction with bluer areas in this region consistent with younger ($\lesssim$500 Myr) populations.  This color map is consistent with the WFPC2 \FV--\FI~color map in Figure 2 of S10 (which was not cross-convolved).

Next, we match up the STIS slit to the WFC3 color images to enable the comparison of the spectroscopic~\ml s with the broad-band colors.  The images were aligned with the STIS slit as described in Section~\ref{ssec:astrometry}.  The white lines in Figure~\ref{colormap} show the location of the STIS slit on the \FB--\FI~color map.  Over the region where spectroscopic~\ml~values have been determined, the \FB--\FI~color varies by more than 1 mag due to dust and stellar population variations.  To account for the spatial variation of \ml~due to both stellar population and dust extinction, we use the effective \ml~(\mleff) that includes dust extinction using the prescription of \citet{Charlot00} and the best-fit $\tau_V$ from our models.  The $I$ band \mleff~along the STIS slit is shown in the bottom panel of Figure~\ref{nucleus_color_m2l_vary}, while the cross-convolved \FB--\FI~color along the slit is shown in the top panel.  The \mleff~is highest in the reddened regions on the eastern side of the nucleus.  Both the color and \mleff~show a minimum at the center, likely due to the contribution of an AGN continuum component.

Figure~\ref{color_m2leff_correlation} shows the correlation of the WFC3 \FB--\FI~color map with the spectroscopic~\mleff~values. We fit this correlation to get a color--\ml~relation; the best-fit is the red solid line.  The idea behind this approach is the same as in \citet{Bell01}, \citet{Bell03}, and \citet{Zibetti09}, but appropriate to the exact populations present in the NGC~404 nucleus as determined using the STIS population synthesis fits.  We fit a linear relation between the logarithm effective spectral~\mleff~and the \FB--\FI~color. The error on this relation is determined in two ways: first, we propagate the errors on the spectroscopic~\ml~determined above, and second, we determine bootstrap errors of the fit by refitting using random sampling with replacement. The first method yields significantly larger errors than our bootstrapped errors, which are minimal. We calculate the 1$\sigma$ uncertainty on our color--\ml~relation from our total error budget, and show these as the thin blue (steeper) and yellow (shallower) lines in Figure~\ref{color_m2leff_correlation}, respectively. We will examine the effect of our color--\ml~relation uncertainties on the dynamical models in Section~\ref{sec:jeans}.. 

\begin{figure}[h]
     \centering
      \epsscale{1.2}
          \plotone{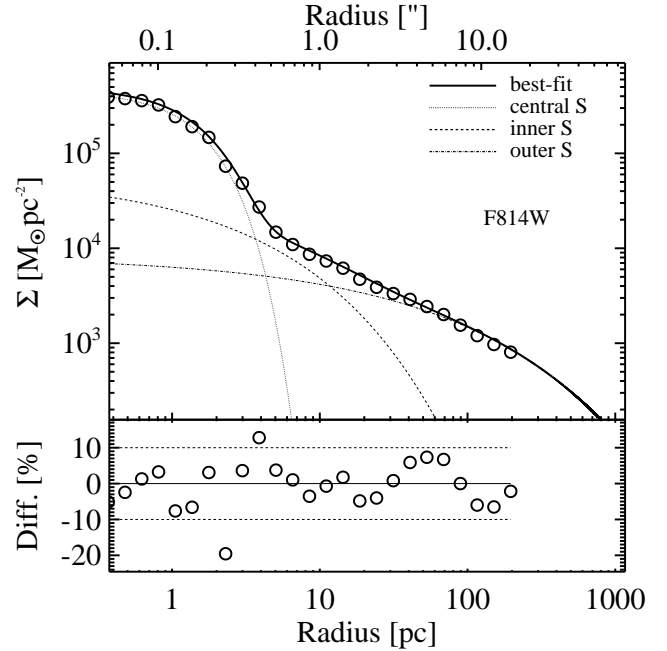}
          \caption[SCP06C1]{Radial mass surface density profile for NGC~404 based on our mass map.  {\em Top panel:} the radial mass surface density profile (open circles), while the solid line shows the \textsc{galfit} radial mass surface density model profile including three single S\'ersic components (see Table~\ref{tab_sersic}). The radial mass surface density profiles for both data and model are extracted using the \textsc{iraf} task \texttt{ellipse}.  {\em Bottom panel:} the percentage difference between the radial mass surface density profile and its \textsc{galfit} model.}
\label{gal_mass_map_comps}
\end{figure}
\begin{figure}[h]
     \centering
      \epsscale{1.2}
          \plotone{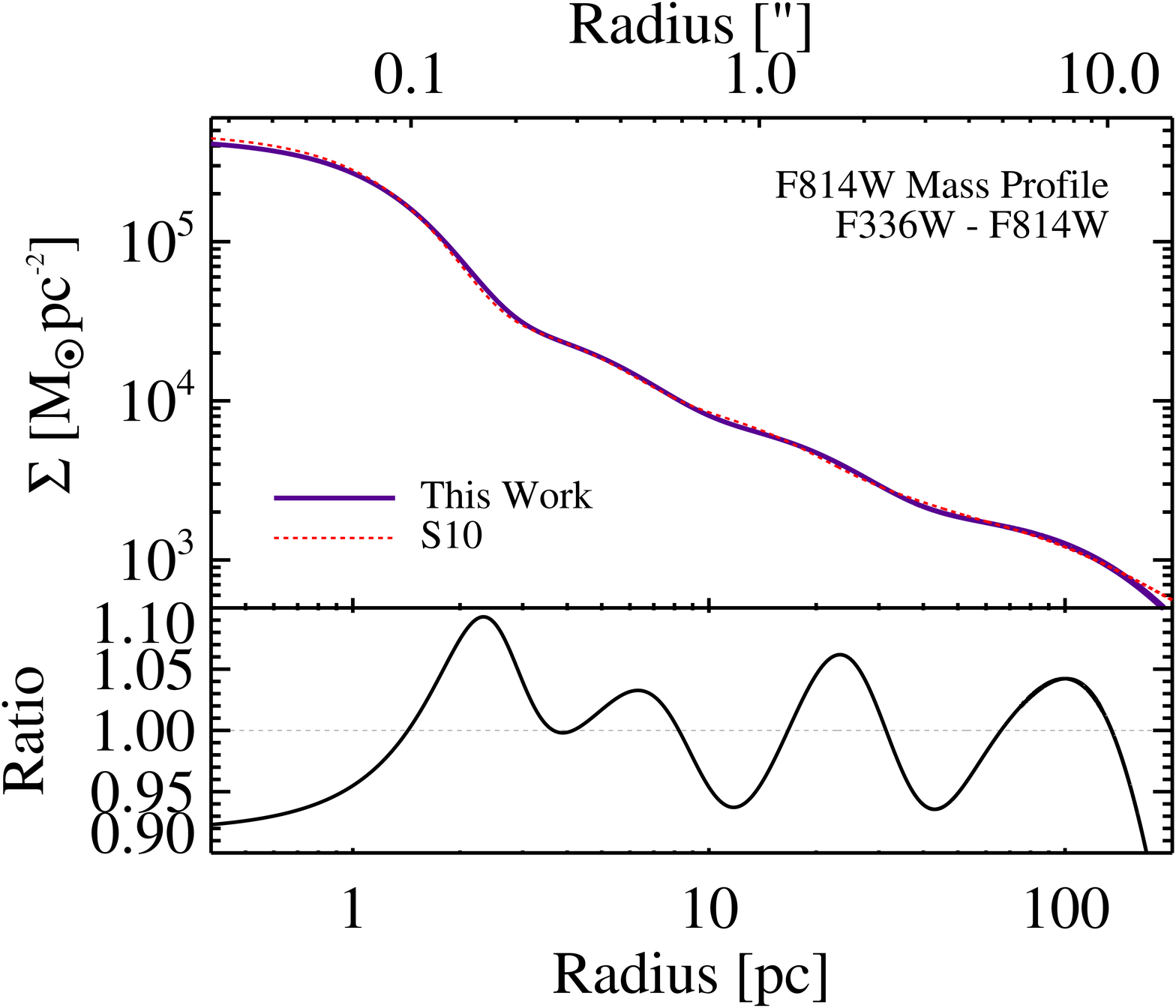}
     \caption[SCP06C1]{{\em Top panel:} comparison of the final MGE model constructed from our nuclear mass map multiplied by our best-fit mass-scaling factor ($\gamma=0.890$, purple line; see Section~\ref{sec:jeans} for details) to that of S10 (red dash line) who assumed a constant~\ml; the mass is obtained by multiplying by the S10 light MGE by their best-fit~\ml~of 0.70. {\em Bottom panel:} ratio of the two mass maps.}
\label{mge_mass_1dim}
\end{figure}

Figure~\ref{color_m2leff_correlation} also compares our best-fit color--\ml~relation to the one predicted by \citet{Bell03} for the Sloan $i$ band based on the Sloan $u-i$ color.  Even accounting for filter transformations, the slope of the \citet{Bell03} relation (purple thick solid thin line) is significantly flatter than our best-fit relation and predicts a much higher mass than our derived relation.  We also plot a color--\ml~relation calculated in our filters by \citet[][green line]{Roediger15}, also using BC03 models with a Chabrier IMF.  This relation has a similar normalization (as expected given the similarities in the models used), but is significantly steeper than our relation.  The differences in slope play the most critical role in affecting our dynamical models, as the overall normalization is scaled to fit the kinematics.  The differences in slope of our relation from the \citet{Roediger15} and the \citet{Bell03} relation is likely due to the unusual SFH in the NGC~404 nucleus, and suggests that knowing the local SFH provides us with important information in mapping out the stellar mass distribution in the nucleus.

As we discussed in Section~\ref{sssec:agn_ssp}, the fraction of young stars in our fits near the center of the galaxy depends significantly on the allowed contribution of the AGN ($\sim$17\%) to the spectrum.   To quantify the effect this has on our \ml, we re-fit the color--\ml~relation using the non-AGN fits for the central spectral bins (thin black line in Figure~\ref{color_m2leff_correlation}).   The \ml$^{\star}$s from these fits are on average 7\% larger than fits including the AGN (column 17 in Table~\ref{tab_bin_ssp}), and this leads to a slightly shallower slope in the color--\ml~relation, but the difference is within our 1$\sigma$ uncertainties.
 
\subsection{Creating a Mass Map and Mass Model}\label{ssec:massmap}    

We describe the creation of a mass map and model for the NGC~404 nucleus in this section.  The first step in this process is to produce an~\mleff~map for the nucleus by applying~\mleff--color relation to the color map.  The \mleff~map in \FI~is shown in the left panel of Figure~\ref{m2l_mass_map}.  This was created using the fitted correlation of the \FI~\mleff~vs. \FB--\FI.

It is straightforward to obtain the mass map for the NGC~404 nucleus by multiplying the~\mleff~map with the \FI~luminosity map pixel by pixel. The right panel of Figure~\ref{m2l_mass_map} shows the WFC3 \FI~mass map of the nucleus of NGC~404. Because the mass map is a weighted version of the \FI~luminosity map, its PSF should be well described by the \FI~PSF.  Comparison of the white (mass map) contours with the red (luminosity) contours shows that the mass distribution in the nucleus is more symmetric than the \FI~light emission profile.   This is because the~\mleff~map is corrected for the dust extinction on the eastern side of the nucleus. We note that the central~\mleff~value from spectroscopy is very close to the best-fit relation, and thus we make no special correction to the map due to the presence of the AGN.  The mass profile of the central regions of NGC~404 is shown in Figure~\ref{gal_mass_map_comps}.

To create a mass model that can be used in dynamical modeling, we need to deconvolve the effects of the PSF on the mass map.  We do this using a \textsc{galfit} model with three S\'ersic components, the inner two to fit the complicated NSC, and the outer to fit the galaxy bulge. Because our mass map extends out to only 25$\arcsec$, we constrain the shape of the outermost bulge S\'ersic component to the values derived by S10: S\'ersic index n$_3$ = 2.5, r$_{\rm eff, 3}$ = 45$\arcsec$ (675~pc), and position angle P.A. = 80\deg, while leaving the normalization and flattening of this component as free parameters.  These parameters are then fit to the 2--D mass map along with all parameters of the inner two S\'ersic profiles using \textsc{galfit}.  The best-fit parameters along with error estimates propagated from the \ml~errors of color--\ml~relation are shown in Table~\ref{tab_sersic}.
The mass map, mass model, and residuals are shown in Figure~\ref{galfit_mass_map}.  The contours of mass density in each panel show both data (black) and model (white) at the same densities to highlight the regions of agreement and disagreement between data and model.  These show that our mass model is more symmetric than our luminosity model.  Our color--\ml~relation has yielded a fairly symmetric mass distribution by up-weighting the \ml~in the dusty regions East (left) of the nucleus, and down-weighting the \ml~where there are young stellar populations to the West (right) of the nucleus.  This increased symmetry shows the success of our color--\ml~relation.

Figure~\ref{gal_mass_map_comps} shows the 1--D radial mass profile of the mass map vs.~the \textsc{galfit} model with the three components shown individually; the residuals in the bottom panel are $<$10\% over the inner region.  This is similar to the level of residuals in the surface brightness fit presented in S10.  Both model and data profiles were determined using the \textsc{iraf} task \texttt{ellipse} \citep{Jedrzejewski87}.

To incorporate this mass model into our dynamical modeling, we need to create a Multi-Gaussian Expansion (MGE) of the model \citep{Emsellem94a}.  We do this by decomposing the \textsc{galfit} model into individual MGEs using the \textsc{mge\_fit\_1d} code by \citet{Cappellari02}; the result is shown in Table~\ref{tab_mge}.  We obtain very similar results by parametrizing the PSF using an MGE and then directly fitting the 2--D mass map, but prefer fitting the \textsc{galfit} model because of the superior handling of the PSF (it allows one to input the actual PSF as a 2--D image).  We fit our MGE out to $\sim$25\arcsec.

We compare our mass MGE to the mass MGE created in S10 from the surface brightness profile without any~\ml~variations in Figure~\ref{mge_mass_1dim}. The two mass profiles are consistent with each other to within 7\% between 0$\farcs$27--25$\farcs$00 (4--375 pc).  However, at the very center, the new model is about 10\% lighter than the original best-fit mass model.

To propagate the error on the color--\ml~relation into our mass maps, we also create mass maps using the 1$\sigma$ uncertainties on the color--\ml~relation.  We also created mass maps from other color--\ml~relations for other filters (i.e.,~in \FV~and \FB).  In general, we find similar results from all maps except for those that we calculate an \FB~\ml. In these cases, excess \FB~light at the center is not fully compensated for by the color--\ml~relations and excess mass is inferred due to the AGN (see column 7 of Table~\ref{tab_jam}).We will discuss the effects these uncertainties have on our dynamical models in Section~\ref{ssec:systemic_uncertainty}.

\section{STELLAR-DYNAMICAL MODELING}\label{sec:jeans}

\begin{table}[h]
\caption{MGE of the \textsc{galfit} Mass Density Map Model}
\centering
\begin{tabular}{cccc}
\hline \hline 
 $j$ &     $I_0$          & $\sigma^{\prime}$ &  $q^{\prime}$\\[0.1cm] 
     &[\Msun/${\rm pc^2}$]&  [$\arcsec$]     &             \\[0.1cm] 
 (1) &       (2)          &      (3)         & (4)         \\[0.1cm] 
\hline
1    &   303071.          &   0.048          &   0.974     \\[0.1cm] 
2    &   258200.          &   0.051          &   0.974     \\[0.1cm] 
3    &     7525.6         &   0.073          &   0.974     \\[0.1cm] 
4    &     4395.6         &   0.084          &   0.974     \\[0.1cm] 
5    &     2643.0         &   0.121          &   0.974     \\[0.1cm] 
6    &     2632.1         &   0.191          &   0.974     \\[0.1cm] 
7    &    16183.5         &   0.251          &   0.974     \\[0.1cm] 
8    &     4477.1         &   0.310          &   0.958     \\[0.1cm] 
9    &     3432.3         &   0.940          &   0.958     \\[0.1cm] 
10   &     1722.6         &   1.039          &   0.958     \\[0.1cm] 
11   &     1353.21        &   1.372          &   0.958     \\[0.1cm] 
12   &      948.75        &   5.265          &   0.997     \\[0.1cm] 
13   &      506.594       &   9.919          &   0.997     \\[0.1cm] 
14   &      204.936       &   25.019         &   0.997     \\[0.1cm] 
\hline 
\end{tabular}
\tablenotemark{}  
\tablecomments{Multi-Gaussian Expansion fit for our best-fit model, created from a \textsc{galfit} model of the mass map shown in Figure~\ref{galfit_mass_map}. Column 1: component number. Column 2: the central surface density (Gaussian amplitude). Column 3: the Gaussian width (in arcseconds) along the major axis. Column 4: the axial ratio.} 
\label{tab_mge}
\end{table}

In this section, we present Jeans modeling of the central BH mass of NGC~404 using the CO bandhead stellar kinematics (Section~\ref{ssec:nifs}) and incorporating the mass surface density map developed in the previous section. 

\subsection{Stellar Kinematics}\label{ssec:stellar_kinematic}

\begin{figure}[h]
     \centering
      \epsscale{1.2}
          \plotone{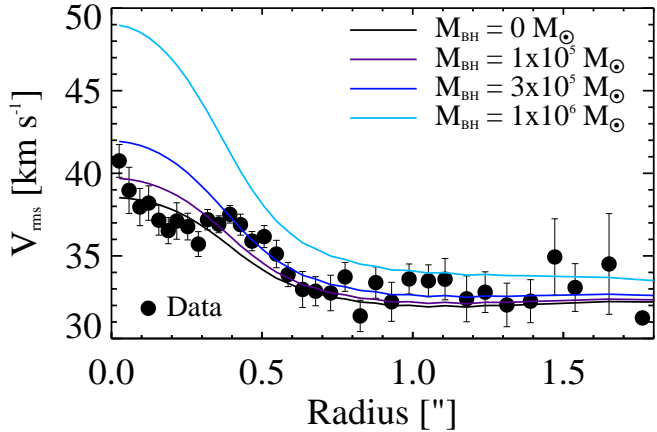}
          \caption{1--D $V_{\text{rms}}$ vs. JAM prediction of BH mass models.  All models are fixed to the best-fit anisotropy parameter ($\beta_z=0.05$), mass scaling factor ($\gamma=0.890$), and inclination angle ($i=20$\deg).  The data are binned radially.  The models show the JAM model predictions at a range of BH masses; the data clearly favors a low BH mass $\lesssim$$10^5$\Msun.}
\label{jam_rms_mbh_models}
\end{figure}

We use the Jeans Anisotropic Modelling (JAM) method and \texttt{IDL} software\footnote{Available from http://purl.org/cappellari/software} as the dynamical model to calculate the mass of central BH, \Mbh. The model relies on an axisymmetric solution of the Jeans equations incorporating orbital anisotropy \citep{Cappellari08}. The anisotropy is characterized by a single anisotropy parameter: $\beta_z=1-\sigma_z^2/\sigma_R^2$, where $\sigma_R$ and $\sigma_z$ are the velocity dispersion in the radial direction and z-direction in ellipsoid aligned cylindrical coordinates. The model calculates the predicted second velocity moment, $V_{\rm rms}=\sqrt{V^2+\sigma^2}$, where $V$ is the mean stellar velocity and $\sigma$ is the velocity dispersion based on an MGE model.  We note we are using a mass model that differs from the luminosity profile of the galaxy; this difference is incorporated into the JAM model by using separate luminosity and mass profile MGEs; however both are axisymmetric and thus may not fully capture the variations in kinematics due to e.g.,~dusty regions.  We also note that because of the proximity of NGC~404, the kinematic maps have some contribution from partially resolved stars that makes the dynamical maps appear less smooth, especially at larger radii. The model has four parameters: (1) the mass of a central BH, \Mbh, (2) the mass scaling factor $\gamma$; while for a normal JAM fit, this would be the dynamical \ml, in our case this parameter is the ratio between the dynamical \ml~and the \ml~predicted by our stellar population fitting (which assumes a Chabrier IMF), (3) the anisotropy, $\beta_z$, and (4) the inclination angle, $i$.

{The models were run over a grid in these parameters, the \Mbh~range is $3\times10^4$\Msun--$3\times10^6$\Msun~and is gridded in step of $\Delta\log$\Mbh = 0.1,
$\gamma$ is gridded in steps of 0.015 between 0.50 \& 2.25, $\beta_z$ ranges from -1 to +1 with a step of 0.05, and the inclination runs from the minimum value allowed by our MGEs, $i=17$\deg~up to 31.5\deg~in step of 0.5\deg.

\begin{figure*}[!htb]
\minipage{0.33\textwidth} 
  \includegraphics[width=\linewidth]{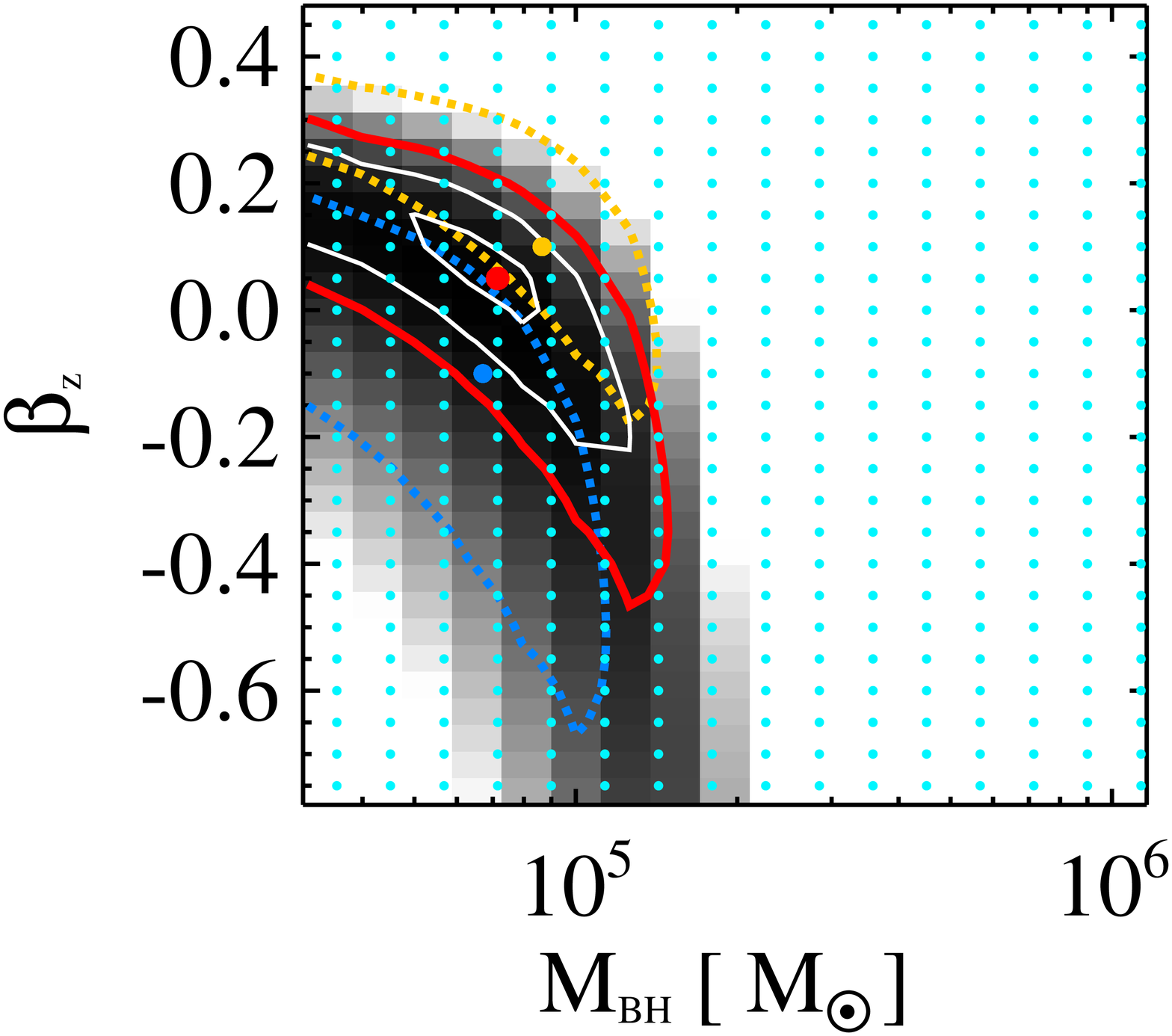}\label{jam_betaz_mbhs}
\endminipage\hfill
\minipage{0.33\textwidth}
  \includegraphics[width=\linewidth]{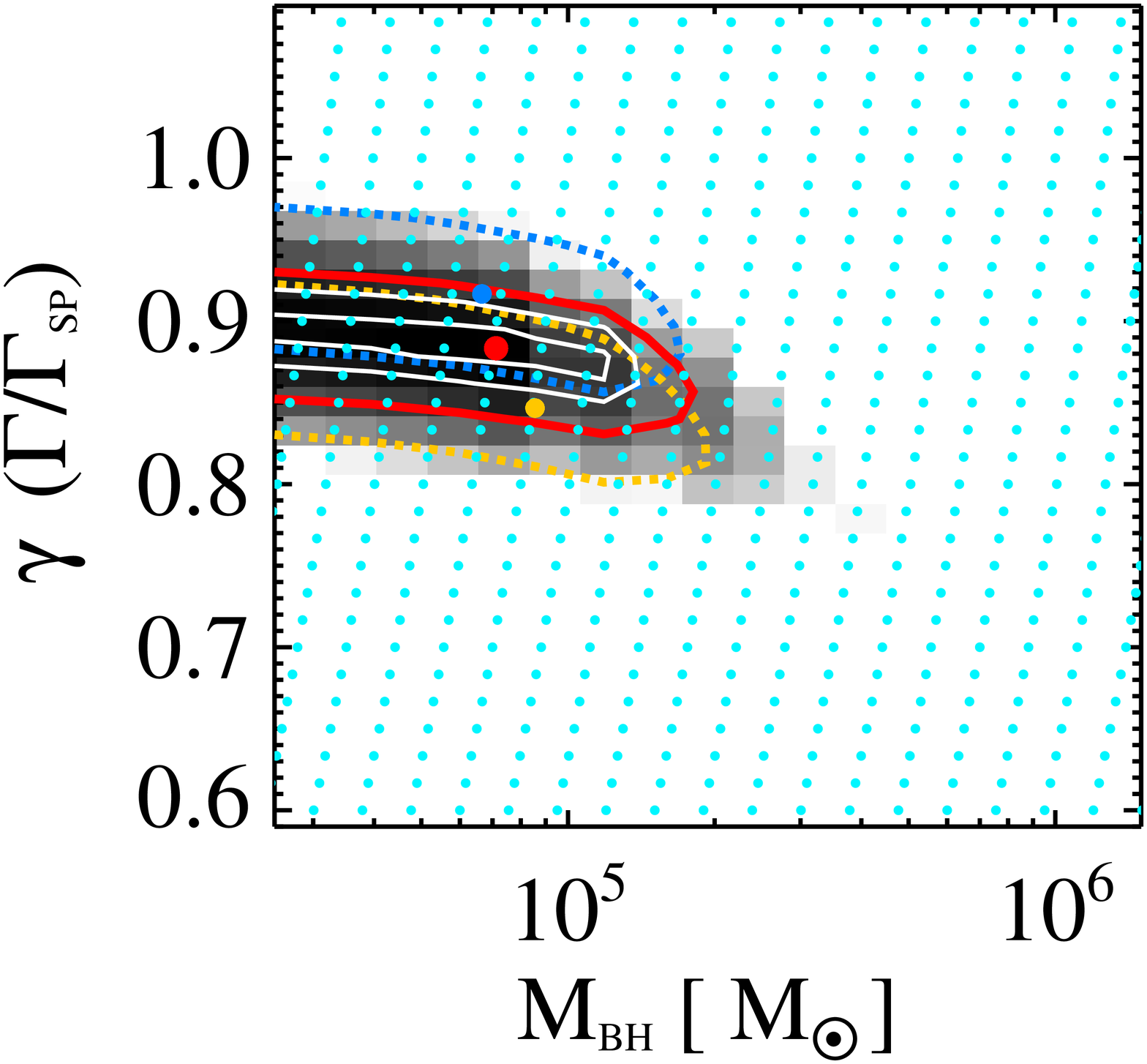}\label{jam_scale_mbhs}
\endminipage\hfill
\minipage{0.33\textwidth}
  \includegraphics[width=\linewidth]{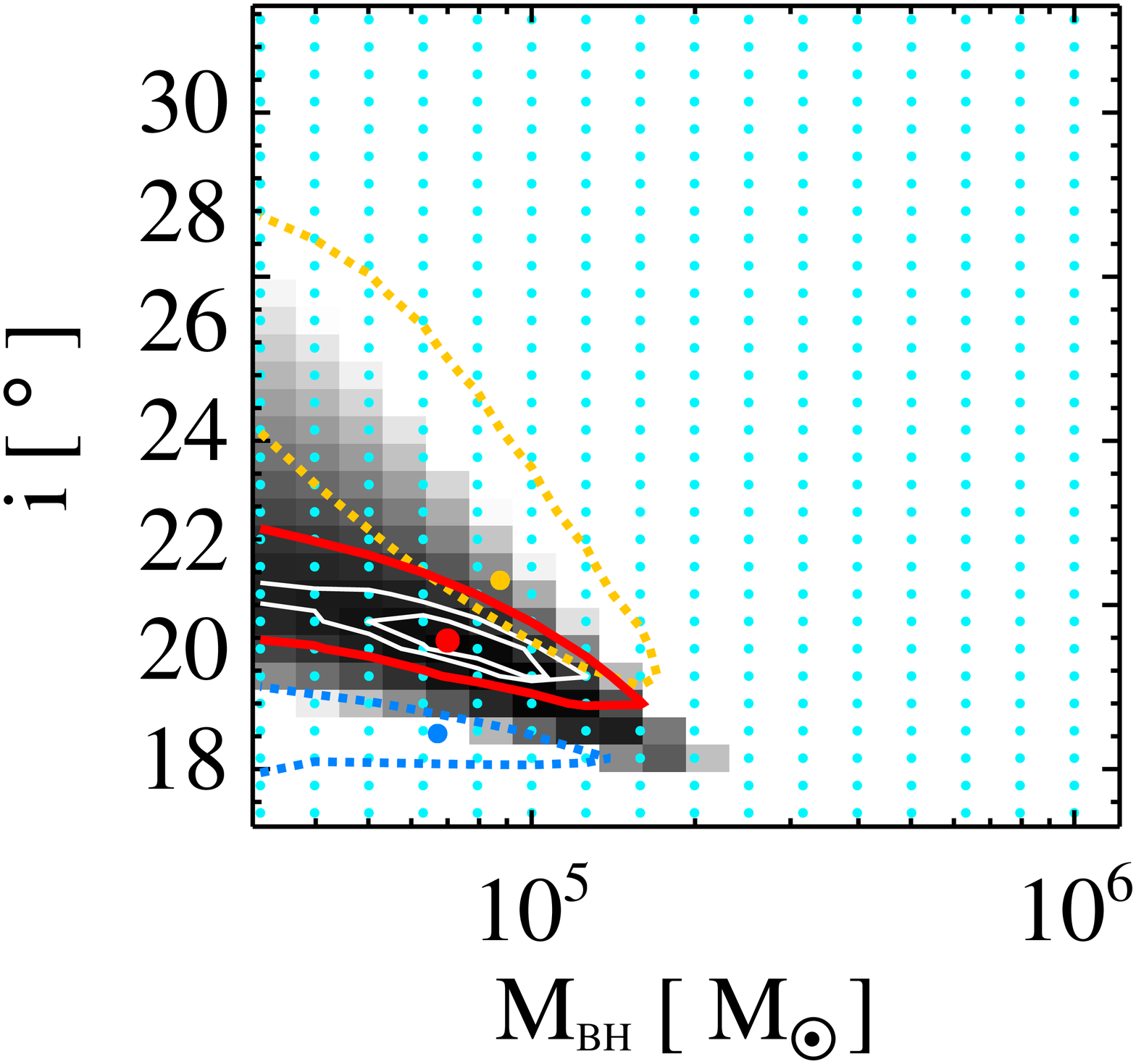}\label{jam_incl_mbhs}
\endminipage
\caption[SCP06C1]{ Best-fit JAM models using the Gemini/NIFS $V_{\rm rms}$ data and our updated mass map MGE model. The models optimize the four parameters \Mbh, $\beta_z$, $\gamma$ (the ratio between the dynamical mass and the stellar population mass), and $i$.  The cyan dots show the grid of models in  each panel. The grayscale shows the likelihood of the best-fit model.  {\em Left panel:} the best-fit anisotropy ($\beta_z$) vs.~M$_{\rm BH}$. The red dot shows the minimum $\chi^2$.  $\chi^2$ contours are shown at $\Delta\chi^2 = 2.30, 6.18$ (white solid lines), ${\rm and \;} 11.83$ (red solid line) corresponding to 1$\sigma$, 2$\sigma$, and 3$\sigma$ confidence levels for two parameters after marginalizing over the other two parameters.   Similarly, the blue and yellow dashed lines illustrate the $\chi^2$ contours of the 3$\sigma$ confidence levels for the mass maps created from the steeper and shallower color--\ml~slope mass map models (corresponding to the thin blue and yellow lines show in Figure~\ref{color_m2leff_correlation}). {\em Middle panel:} the best-fit mass scaling factor, $\gamma$ vs.~\Mbh; markings as in left panel. {\em Right panel:} the inclination angle $i$ of the galaxy vs.~\Mbh.}
\label{jam_models}
\end{figure*}  


We compute the models on a grid of $\beta_z$, \Mbh, $\gamma$, and $i$. At each grid point we evaluate the $\chi^2$ of the the predicted $V_{\rm rms}$ relative to the kinematic data. We run the dynamical modeling to fit these four parameters ($\beta_z$, \Mbh, $\gamma$, $i$) in 2--D using 920 kinematic data points (degrees of freedom, dof) within a radius of $\sim1\farcs5$ of the center. Figure~\ref{jam_models} shows $\chi^2$ contours as a function of \Mbh~vs. the anisotropy parameter $\beta_z$ (left), scaling factor $\gamma$ (middle), and inclination angle $i$ (right) after marginalizing over the other two parameters.  The minimum reduced $\chi^2_r\sim1.26$~is found at \Mbh~= $7.0\times10^4$\Msun, $\beta_z\sim0.05$ (i.e., nearly isotropic), $\gamma=0.890$, and $i=20$\deg.  The solid contours show the  $1\sigma$, $2\sigma$ (white), and $3\sigma$ (red) levels (or $\Delta\chi^2=2.30$, $6.18$, and $11.83$) for two parameters; we choose the confidence limits for two parameters to accurately capture the uncertainties shown in our two parameter plots above after marginalizing over the third and the fourth parameter in each plot.  The BH is only detected at the 1$\sigma$ level.   Given the restrictions JAM models place on the orbital freedom of the system (relative to e.g., Schwarzschild models), it is common to use 3$\sigma$ limits in quoting BH masses (see Section~\ref{sssec:models} for more details).  Therefore, we use the 3$\sigma$ upper limit of \Mbh~$<1.5\times10^5$\Msun.  This is similar to (but slightly larger than) the upper limit of the light-follows-mass model in S10 (using a $\Delta\chi^2 = 9$ limit).

Figure~\ref{jam_rms_mbh_models} shows the 1--D $V_{\rm rms}$ vs. JAM predictions for our axisymmetric mass model with black holes with different masses. The $V_{\rm rms}$ values shown are bi-weight averaged over circular annuli.  These data are compared with lines for different BHs; these show that BHs with \Mbh$\gtrsim10^5$\Msun~do not provide a good fit to the data.  Particularly, the innermost bin looks consistent (within the errors) with the $3\times10^5$\Msun~model. Beyond $\sim$0$\farcs$3, the data look consistent with anything \textless$3\times10^5$\Msun, and beyond $\sim$0$\farcs$7 there is little difference between the models.

The most significant change in comparison with S10 is the anisotropy parameter, $\beta_z$.  Our best model is more isotropic ($\beta_z=0.05^{+0.05}_{-0.15}$) than the best fit of $\beta_z=0.5$ found in S10.  Given the observed isotropy in other galaxy nuclei \citep[e.g.,][]{Schodel09, Seth14}, we interpret this as a sign of the success of our mass model (which incorporates variations in~\mleff) correctly representing the mass distribution in this system.

This analysis does not include any gas mass in the nucleus; we show here that we expect this assumption to have a minimal effect.  \ion{H}{1} gas is present at large radii in NGC~404 but has a hole at the center \citep{delrio04}.  There is CO emission from the regions in and around the nucleus: assuming a distance of $D=3.06$~Mpc, the total molecular gas mass is estimated to be $6\times10^6$\Msun~\citep{Wiklind90}.  Most of this gas is to the NE of the nucleus, and only a few percent is within the central 10$\arcsec$, thus the amount within the region we are modeling is $\lesssim$$10^5$\Msun.  This gas mass is less than 1\% of the stellar mass in the region we are modeling the kinematics, and based on the molecular hydrogen emission map (Section~\ref{sec:gasmodeling}), this gas is likely spread widely across the nucleus; thus we do not expect this component to affect our \ml~or BH mass estimates.  We examine the kinematics from the warm molecular hydrogen emission in Section~\ref{sec:gasmodeling}.

\subsection{Uncertainties due to mass models}\label{ssec:massmodel_uncertainty}

The confidence intervals in our analysis above are based on the kinematic measurement errors and do not include any systematic uncertainties in the mass model. In this section, we examine the mass model uncertainties by analyzing additional, independent mass model images.

The primary uncertainty in \Mbh~comes from the central stellar mass profile.  We can get a rough uncertainty on the BH mass by examining the uncertainty in the total mass in the central pixel of our mass map. The central pixel in our mass map has a projected stellar mass of 5.1$\times10^5$\Msun~(with an area of 1.8~pc$^2$); the dominant uncertainty in this photometric estimate of the central mass is the uncertainty in the central \mleff. This uncertainty is 20\% or $\sim$$1\times10^5$\Msun.   We can expect our BH mass uncertainty to be comparable to the uncertainty in the mass in this central pixel.   We now examine the systematic uncertainty due to our mass map using several methods.  Overall, we find that the systematic errors on our BH mass are all consistent with the 3$\sigma$ upper limit of $1.5\times10^5$\Msun. 

\subsubsection{Color--M/L Relation Errors}

We calculated the 1$\sigma$ uncertainties on our color--\ml~relation, shown as blue and yellow lines in Figure~\ref{color_m2leff_correlation}.  We created mass maps and MGEs from these 1$\sigma$ steeper and shallower color--\ml~relations and ran a full grid of JAM models for both.  Figure~\ref{jam_models} show the resulting 3$\sigma$ confidence levels. The effect on the BH mass from these models is minimal, probably because even with the errors, the color--\ml~relations are quite similar at the bluer colors found near the nucleus.  However, the larger changes in $\gamma$ and $i$ show that these parameters are quite sensitive to the assumed color--\ml~relation, and that our formal errors do not capture the full extent of the uncertainty in these parameters. 

As noted, the 1$\sigma$ errors on our relation still do not encompass the color--\ml~relations derived in previous work.  To test the effect of these more extreme color--\ml~relations, we also derived the stellar mass map based on \citet{Bell03} relation (the purple thick solid line in Figure~\ref{color_m2leff_correlation}) to examine how much the slope in the color--\ml~relation affects the constraints on the BH mass.  The resulting best-fit BH mass is lower, \Mbh$_{\rm, Bell03}$~= $3.3\times10^4$\Msun, but the most notable difference was the mass scaling factor of $\gamma_{\rm Bell03}=0.33$; this factor of $\sim$3 is expected due to the higher normalization of the \citet{Bell03} relation.   

We also fit our source using the color--\ml~relation derived without including an AGN component in the fit.  As noted in Section~3.3,~
the central \ml$^{\star}$~and NSC mass increase by 7\%, which suggests we will derive a less massive BH at the center of NGC~404.  Using this non-AGN color--\ml~relation to create our mass model, we get JAM models with a best-fit BH mass of \Mbh~$=~$6.3$\times10^4$\Msun~and $\gamma=0.95$ fixing the anisotropy to $\beta_z=0.05$ and the inclination $i=20$\deg.  Therefore, the assumption of an AGN contribution has minimal effect on our results, with a best-fit BH mass 9\% higher than the non-AGN case.

To summarize, the errors on our color--\ml~relation suggest minimal changes in our best-fit BH mass, and all reasonable models are consistent with our 3$\sigma$ BH mass upper limit of $1.5\times10^5$\Msun.  However, substantial changes in the best-fit $\gamma$ and inclination are seen using different color--\ml~relations.

\subsubsection{Mass Maps from Other Filters}

Our default mass model is created using the \FB -- \FI~color map and the \FI~image.  As another way of gauging the uncertainty in the central mass profile, we also created mass maps using alternative wavelength images and~\mleff~vs.~color relations.   We then create new MGEs from \textsc{galfit} model fits to these maps, and repeat the JAM modeling.  We work on three color maps of \FB -- \FI, \FV -- \FI, and \FB -- \FV, then create a STIS~\mleff~vs.~color relation as described in Section~3.3,~
For each color, we create a~\mleff~map and mass map for the \FB, \FV, and \FI~filter.    Therefore, in total, we create nine WFC3 mass maps and their corresponding \textsc{galfit} model mass maps.  We present the best-fit JAM models generated from these mass maps in Table~\ref{tab_jam}.   

When using the \FV~and \FI~images as the basis for our mass maps, our results are fully consistent with the results we obtain for our default assumptions, suggesting that our mass model uncertainties are lower than the uncertainties from the kinematic fitting.  However, when using the \FB~images to make the maps, the excess emission at the center increases the stellar mass density and decreases the best-fit \Mbh.  This effect is likely due to AGN emission and is still only at the 2$\sigma$ level.  

To test any possible systematic errors caused by our \textsc{galfit} model, we also directly fit the mass map with a MGE model.  This makes fewer assumptions about the shape of the mass profile but requires that we approximate the PSF as an MGE as well.  This MGE fit results in a best-fit \Mbh~$=8.0\times10^4$\Msun~(14\% higher than the \textsc{galfit} model) with the same $\beta_z$, $\gamma$, and $i$ values.

\subsubsection{IMF variations}

In calculating our stellar population \ml, we assume a Chabrier IMF, and the best-fit $\gamma$ of 0.890 suggests our overall IMF cannot be too different from this.  Furthermore, Figure~\ref{jam_rms_mbh_models} shows that radially, the deviations of $V_{\rm rms}$ from the predicted model are at most $\sim$5\%, suggesting radial variations in the mass scaling factor (which depends on $V_{\rm rms}^2$) are at $\lesssim$10\%.  Nonetheless, variation of the IMF within the nucleus could affect our BH mass estimate.  While there is evidence for IMF variations at the centers of elliptical galaxies \citep{Conroy12,Cappellari12,Martin-Navarro15},  there is little knowledge of IMF variations on the parsec scales probed here.  The only direct information on the IMF in NSCs comes from our own Milky Way, and at times the IMF near the center of the Galaxy has been suggested to be very shallow at the high-mass end relative to standard IMFs \citep[e.g.,][]{Bartko10}.  However, the more recent work of \citet{Lu.R13} based on AO observations of individual stars suggests a modestly shallower IMF for young stars in the Milky Way NSCs of $dN/dM \propto M^{-1.7\pm0.2}$ .  We note that for older stellar populations, both shallower and steeper IMFs increase the \ml~\citep[e.g.,][]{Cappellari12}.  Given the low inferred BH mass, there is little room for such an IMF at the center of NGC~404.

\subsection{Additional Sources of Uncertainty}\label{ssec:systemic_uncertainty}

In this section, we discuss additional sources of uncertainties due to our dynamical modeling techniques and the assumed center.

\subsubsection{Dynamical Modeling Uncertainties}\label{sssec:models}

We have chosen to dynamically model our stellar kinematics with Jeans' Anisotropic models \citep[JAM][]{Cappellari08}, which provide good fits to real ETGs on large scales \citep{Cappellari16Rview}. These models make strong assumptions on the form of orbital anisotropy and have less orbital freedom relative to Schwarzschild models \citep[e.g.,][]{Vandenbosch08}.  Particularly, JAM models predict only the second moment of the line-of-sight velocity distribution, $V_{\rm rms}$, which is known to provide BH mass constraints that are degenerate with the anisotropy (as seen in Figure~\ref{jam_models}). The JAM models incorporate an anisotropy parameter, but this parameter still imposes strong constraints on the allowed orbits which can make the resulting parameter errors too small.  Furthermore, we have assumed a constant anisotropy.  Despite these shortcomings, our models likely do accurately represent the NSC in NGC~404, as the anisotropy has been found to be small and fairly constant in the Milky Way, M32, and Cen~A \citep{Schodel09,Verolme02,Cappellari09}, as well as the more distant M60-UCD1 \citep{Seth14}.

JAM models give BH mass estimates consistent with Schwarzschild models \citep{Cappellari09,Cappellari10,Seth14,Thater16} and high precision maser disk measurements \citep{Drehmer15}.  Schwarzschild models, which include all physical orbits of stars within given a potential, do have larger, more realistic, uncertainties.  \citet{Seth14} find errors in BH mass from M60-UCD1 from JAM models that are 2-3 times smaller than the errors from the Schwarzschild models. Another example is the BH mass estimate for NGC~1277 based on JAM by \citet{Emsellem13}, where , suggested a significantly lower mass than the published value by \citet{vandenBosch12}. And this JAM lower value was later confirmed with Schwarzschild models based on new integral-field data by \citet{Walsh16}. This shows that Schwarzschild, although in principle better because of its generality, is not always more robust.  Therefore, we have chosen to use a 3$\sigma$ upper limit for the BH mass here, which is also larger than other systematic errors presented above.

\subsubsection{Central Position Uncertainties}\label{sssec:cen_pos} 

We discuss the uncertainties caused by our assumption of the dynamical center position of the galaxy.  The choice of the center can have a large effect on BH mass determinations \citep[e.g.,][]{Jalali12}, and appears to have a strong effect on our gas kinematic models (Section~\ref{sec:gasmodeling}).  We have determined both a kinematic and photometric center; these centers are offset by about a half a pixel, or $\sim$0.4~pc. Our default stellar dynamical model assumes the kinematic center; when we run the JAM model on the photometric center, we get the best fit  BH mass of $\sim$5.6$\times10^4$\Msun. This suggests a $\sim$20\% systematic error, similar to the 1$\sigma$ error bars.  We also test to ensure that our mass map extends radially far enough out; we remake our MGEs from mass maps extending out to different radii and find that the results are stable if the mass map has a radius that is larger than 6$\arcsec$. Using an MGE that does not go out as far results in a systematic increase in the BH mass; this increase is $\sim$30\% if we narrow down the mass map radius to 1$\farcs$5.

To conclude, our constraints on the BH mass appear to be robust with little evidence in our tests for significant systematic errors.  Thus, we present our \Mbh~$<1.5\times10^5$\Msun~as a robust 3$\sigma$ upper limit.

\begin{table*}[ht]
\caption{JAM-Stellar Dynamical Modeling Results in Different Mass Map Filters and Colors}
\centering
\begin{tabular}{cccccccccc} 
\hline 
$j$  & Color                     &Filter&      Map      &    \Mbh         &       $\beta_z$     &        $\gamma$       &         $i$      \\
     &                           &      &               &  $10^4$~[\Msun] &                     &                       &        [\deg]    \\
(1)  &    (2)                    & (3)  &      (4)      &       (5)       &        (6)          &        (7)            &         (8)      \\[1mm]
\hline
 1   &\multirow{3}{*}{\FB -- \FI}& \FB  &\textsc{galfit}&$4.0^{+3.5}_{-2.5}$&$0.15^{+0.10}_{-0.10}$ &$1.895^{+0.045}_{-0.105}$&$23.0^{+2.5}_{-2.0}$\\[1mm]
 2   &                           & \FV  &\textsc{galfit}&$6.5^{+1.5}_{-0.8}$&$0.00^{+0.05}_{-0.10}$ &$0.950^{+0.055}_{-0.090}$&$21.0^{+1.5}_{-1.0}$\\[1mm]
 3   &                           & \FI  &\textsc{galfit}&$7.0^{+1.7}_{-0.4}$&$0.05^{+0.05}_{-0.15}$ &$0.890^{+0.060}_{-0.045}$&$20.5^{+1.0}_{-2.0}$\\[1mm] 
\hline
 4   &\multirow{3}{*}{\FB -- \FV}& \FB  &\textsc{galfit}&$3.7^{+3.1}_{-2.3}$&$0.10^{+0.15}_{-0.10}$ &$1.970^{+0.075}_{-0.120}$&$23.0^{+2.0}_{-2.5}$\\[1mm] 
 5   &                           & \FV  &\textsc{galfit}&$7.2^{+1.6}_{-0.7}$&$-0.05^{+0.10}_{-0.10}$&$0.935^{+0.090}_{-0.105}$&$21.5^{+1.5}_{-1.5}$\\[1mm] 
 6   &                           & \FI  &\textsc{galfit}&$6.8^{+1.7}_{-0.8}$&$0.05^{+0.05}_{-0.05}$ &$0.905^{+0.075}_{-0.090}$&$19.5^{+1.0}_{-2.0}$\\[1mm] 
\hline
 7   &\multirow{3}{*}{\FV -- \FI}& \FB  &  Direct         &$4.3^{+3.4}_{-2.3}$&$0.10^{+0.10}_{-0.15}$ &$1.910^{+0.105}_{-0.120}$&$22.5^{+2.5}_{-1.5}$\\[1mm]
 8   &                           & \FV  &  Direct         &$7.5^{+2.0}_{-0.9}$&$-0.05^{+0.05}_{-0.05}$&$0.920^{+0.075}_{-0.090}$&$21.5^{+1.5}_{-2.0}$\\[1mm]
 9   &                           & \FI  &  Direct         &$7.2^{+1.4}_{-0.5}$&$0.05^{+0.05}_{-0.10}$ &$0.860^{+0.045}_{-0.105}$&$20.0^{+2.0}_{-1.5}$\\[1mm]
\hline
\end{tabular}
\tablenotemark{}
\tablecomments{ Exploring the systemic uncertainty in our dynamical models using mass maps created with different filters and colors. Column 1: number of nuclear mass map.  Column 2: the color used to create the color--\ml~relation.   Column 3: the \hst~ filter used in creation of the mass map.  Column 4: Method used to create MGE; Galfit means that mass maps were fitted to a Galfit model which was then fit to an MGE model, ``Direct'' means that the MGE was created directly from the mass map. Column 5, 6, 7, and 8: the best-fit BH mass (\Mbh), anisotropy ($\beta_z$), scaling factor ($\gamma$), and  inclination angle ($i$) of each best-fit model determined from Jeans models. All error bars are determined by propogating the uncertainty in the color--\ml~relation through the dynamical models (e.g., the blue and yellow lines in Figure~\ref{color_m2leff_correlation} and Figure~\ref{jam_models}).} 
\label{tab_jam}
\end{table*}

\section{Gas Kinematics Modeling}\label{sec:gasmodeling} 

To independently constrain the~\ml~of the nucleus and the mass of the central BH, we model the kinematics of the molecular hydrogen gas. A gas dynamical model of NGC 404 was also constructed by S10. Our models use the same data, but we make some different assumptions to test the robustness of the measured BH mass. We find the models are extremely sensitive to our choice of center, and furthermore, that the gas motions near the center suggest the gas is either distant from or not in rotation around the nucleus.  We therefore suggest the NIR emission from molecular hydrogen gas cannot be used to constrain the BH mass in this object.

For the stellar potential, we use the MGE model derived in Section~\ref{ssec:massmap}. As with the JAM models presented above, the deconvolved stellar mass model is multiplied with an additional scaling in~\ml. The BH is modeled as a point source. 

\begin{figure*}
\begin{center}
 \includegraphics[width=1.0\textwidth,clip=]{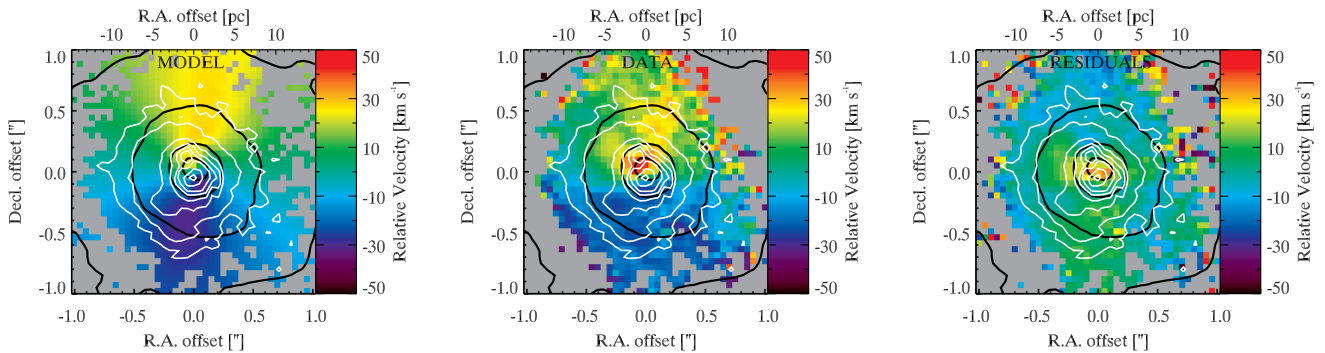}
 \caption{Comparison of the velocity fields of our best-fit kinematic center model (left) with the Gemini/NIFS H$_2$ velocity field (center) and velocity residuals (right). The black contours denote the dust-emission-corrected $K$-band image of the NSC from S10. White contours show the intensity of the H$_2$ gas. Despite being optimized for fitting the central $12\times12$ pixels, the worst residuals are found at the very center of the cluster.}
\label{stellar_photo_center_differ}
\end{center}
\end{figure*}

\subsection{Thin disk tilted ring models}

We model the H$_2$ 1-0 S(1) transition kinematics with a tilted ring model, similar to the modeling approach used in S10. In summary, our models assume that the H$_2$ emitting gas is rotating in thin rings on circular orbits around the center of the nucleus of NGC 404. At each radius, the velocity of the gas traces the enclosed mass in the spherical symmetry approximation \citep[see e.g.,][]{Neumayer07}.

\begin{eqnarray}
v_c^2(r) = \frac{G M(<r)}{r}.
\end{eqnarray}

Each ring of gas can have its own inclination and position angle. We constrain these parameters for rings in the outer parts by fitting ellipses to the velocity field with \textsc{Kinemetry} \citep{Krajnovic06}. Within 0$\farcs$1 of the center of the nucleus, where the PSF is more important than in the outer parts, we optimize for the best values. As was found by S10, there is a kinematic twist close to the center, where the position angle (P.A.) changes by $70^{\circ}$.

For our model we generate a spectral cube with the same spectral sampling as the NIFS IFU, but spatially oversampled by a factor of 10. For each ring, we distribute its flux over the cube according to its spatial distribution and the velocity along the ring to mimic our observations. We also spectrally convolve the flux of each ring by a Gaussian with a width of 20~km~s$^{-1}$ to mimic the intrinsic dispersion of the disk.  We then convolve each spectral slice with the oversampled NIFS PSF and resample the cube in order to compare it with our data.

Contrary to S10, we fit our dynamical models directly to the kinematic data without first symmetrizing them to enable more reliable data versus model comparisons.  We calculate the $\chi^2$ for both the predicted velocity field and the predicted dispersion field. However, since the measured velocity field is more constraining than the dispersion field, we use only the $\chi^2$ of the velocity field to determine our best-fit model. To optimize our fit for the BH mass, we follow S10 in only using the central $12 \times 12$ pixels for our $\chi^2$ determination. Using a wider field for the calculation of the $\chi^2$ gives qualitatively the same results.

S10 modeled the surface brightness of the H$_2$ gas as two exponential disks. In reality, the distribution of the gas is very irregular, as is visible in the H$_2$ intensity contours in Figure~\ref{stellar_photo_center_differ}. We therefore deconvolve the line flux derived from the IFU data of the H$_2$ gas by fitting a number of Gaussians to the peaks in emission close to the center with \textsc{galfit}. This different flux model leads to slightly different gas dynamical models compared to those of S10. However, this improvement in the modeling did only marginally change the best-fit black hole mass.

\begin{figure}
\begin{center}
 \includegraphics[width=0.53\textwidth,clip=]{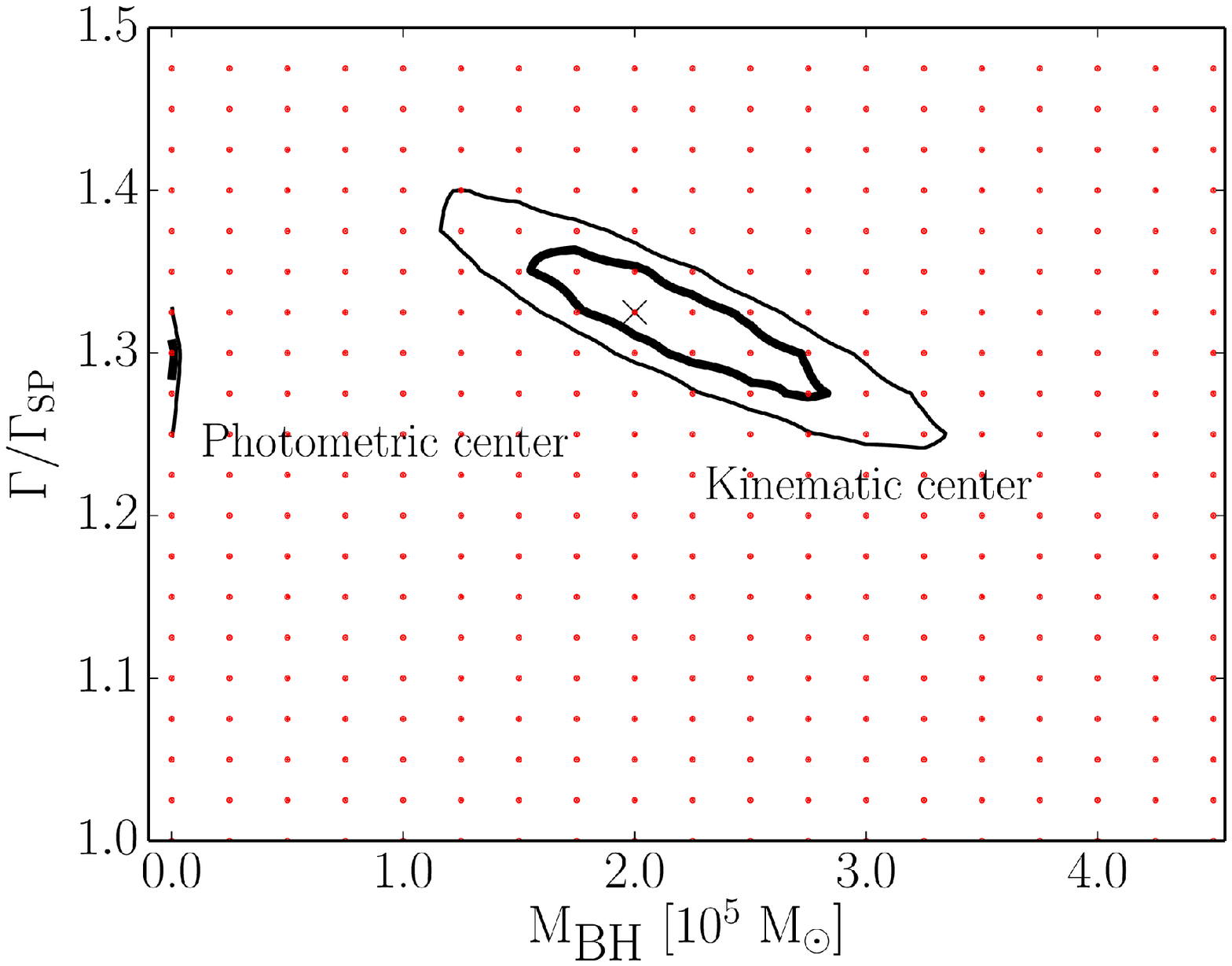} 
\caption{The best fit~\ml~and BH mass for the kinematic center and the photometric center. The~\ml~is expressed as the ratio of the dynamical mass and the the stellar~\ml~found by fitting SSP models to the STIS data (the same $\gamma$ parameter fit using the JAM models). The red dots show our model grid. The $\chi^2$ at each grid point is optimized for the inclination. The contours show the 1$\sigma$ (thick line) and 3$\sigma$ model space allowed by the data.}
\label{gas_dyn_models}
\end{center}
\end{figure}

\subsection{Model grid}

S10 found a best-fit inclination of 37\deg~for the gas disk. We allow the inclination to vary between 27\deg~and 47\deg, in steps of 5\deg.  As with the stellar population models, we include an additional mass scaling factor $\Gamma/\Gamma_\mathrm{SP}$, which was allowed to vary between 1.0 and 1.5 in steps of 0.025.  We used black hole masses between 0--$4.5\times10^5$\Msun~in steps of $2.5\times10^4$\Msun.

The exact location of the center of NGC~404 is uncertain. The photometric center, which was derived in S10 by correcting the Gemini/NIFS continuum emission for the emission of hot dust, is offset by 0.37~pc ($0\farcs025=0.5$~pixels) from the kinematic center. We, therefore, run our grid of models for two different centers.

\subsection{Results}

Figure~\ref{gas_dyn_models} shows the results of the modeling of the H$_2$ 1-0 S(1) gas kinematics. When using the kinematic center, as was done by S10, we derive a BH with mass $\sim$$2.0^{+1.4}_{-0.9}\times10^5$\Msun (3$\sigma$). This is somewhat lower than the gas dynamical mass estimate found in S10, but within the errors of that previous determination.  However, when using the photometric center we find that we cannot dynamically confirm the presence of a BH, i.e., our models are consistent with no BH.  The best-fit mass scaling factor of the cluster is 1.325$\pm$0.075.  This is significantly higher than the mass scaling factor from the stellar dynamical models ($0.890_{-0.060}^{+0.045}$).

The reduced $\chi^2_r$ of our fits are high: $\chi^2_r=109$ for the kinematic center, and $\chi^2_r=164$ for the photometric center. Although it is possible that we have underestimated the uncertainties on the velocity data, the most likely contribution to the high $\chi^2_r$ is systematic; the models simply do not provide a good fit to the data.   In particular, large residuals in velocity are found close to the center of the cluster (Figure~\ref{stellar_photo_center_differ}). It is not possible to remove this (positive) velocity peak by shifting the center. We note that this residual peak was also visible in the residual map  of S10 (see their Figure 16).

Since shifting the center of our models by 0.4 pc (half a pixel) has a huge influence on the $\chi^2_r$ and can either introduce or take away the necessity of an BH in our models, we do not consider the $2\times10^5$\Msun~for the mass of the BH  a significant result, especially since the peak in the dispersion map of the H$_2$ gas does not coincide with either the kinematic or photometric center. Fitting the H$_2$ gas with two disks, one with position angle (P.A.) of 50\deg~and one with a P.A. of $-20\deg$ can reproduce qualitatively some of the features seen in both the velocity map (the double-lobed structure) and the off centered peak seen in the dispersion map. However, we find that it is still not possible to model the positive central peak in velocity. It is likely that part of the observed motion of the H$_2$ gas is therefore not tracing the potential, and might be either due to infalling or outflowing gas, or foreground gas at larger galactocentric radii.  This may also explain the mismatch in the mass scaling factor between the best-fit gas dynamical and stellar dynamical models. 
 
\section{Discussions}\label{sec:dis}

\subsection{The NGC~404 BH}\label{ssec:ngc404bh}

Previous work has already provided strong evidence for the presence of an accreting BH in the nucleus of NGC~404, and our observations of nuclear variability (Section~\ref{ssec:nucleusvary}) and blue continuum in the STIS spectrum (Section~\ref{ssec:ssp}) have added to this evidence. Our dynamical constraints provide a robust upper limit of 1.5$\times$10$^5$\Msun~on this BH.  We note that our upper limit makes the NGC~404 BH consistent with the radio non-detection of a compact core in NGC~404 by \citet{Paragi14}; they place a $\sim$3$\times10^5$\Msun~upper mass limit on the BH based on fundamental plane measurements.

NGC~404 is thus unique; it contains a massive black hole whose mass is dynamically constrained to be $\lesssim$$10^5$\Msun.  This alone can provide some evidence on the mechanism that formed BHs in the early universe \citep[e.g.,][]{Volonteri10, Greene12}.  The dynamical upper limits on possible BHs in M33 and NGC~205 are an order of magnitude lower than for NGC~404 \citep{Gebhardt01, Merritt01a, Valluri05}, however, there is no evidence that BHs exist in either of these objects.  The lowest mass BH with a dynamical measurement is in NGC~4395 \citep[$4\times10^5$\Msun;][]{Denbrok15}.  However, some similar BH estimates do exist for very faint galaxies detected as broad-line AGN \citep{Reines13,Moran14,Reines14}.  The lowest mass BH with any kind of mass estimate was recently presented by \citet{Baldassare15} who find a virial BH mass estimate from the broad H${\alpha}$ line of 5$\times$10$^4$\Msun.

BH mass detections have also been claimed in globular clusters, and these are typically between 10$^{4-5}$\Msun \citep{Gebhardt05, Noyola08, Lutzgendorf11}.  However, these detections remain controversial, and lower mass systems can be explained by clusters of dark remnants at the centers of clusters \citep{Lutzgendorf13b, Denbrok14}.  

\begin{figure}[ht]
     \centering
      \epsscale{1.2}
          \plotone{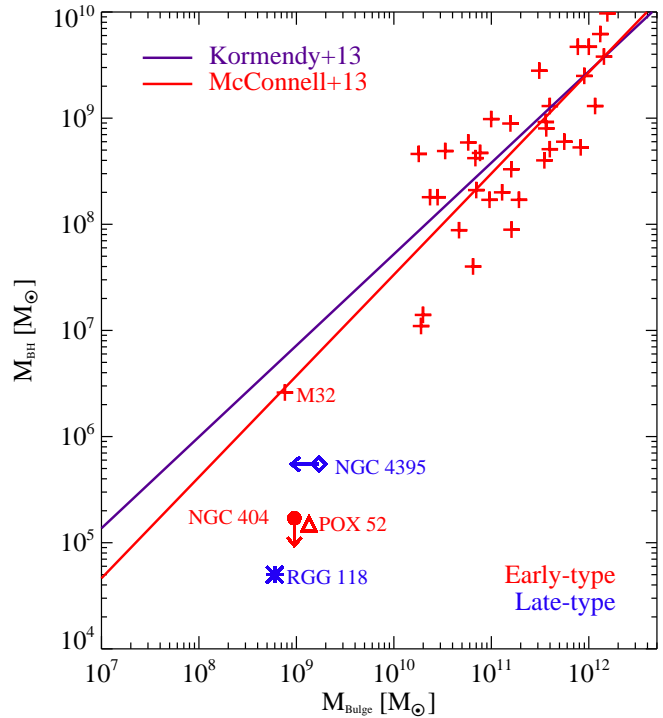}
     \caption[SCP06C1]{ NGC~404 in context of the \Mbh$-M_{\rm Bulge}$ scaling relations for ETGs (red pluses). We note the red color indicates the ETGs, while blue color illustrates the late-type counterparts.  The red and purple lines show the fit from \citet{McConnell13b} and \citet{Kormendy13}, respectively.   The NGC~404 BH is shown as a red dot with a downward arrow at the  3$\sigma$ upper limit; it falls well below the \Mbh$-M_{\rm Bulge}$ scaling relations. We also show the mass determinations of dwarf AGN POX 52 \citep{Barth04,Thornton08} and NGC 4395 \citep{Filippenko03,Denbrok15}, broad H${\alpha}$ emission line AGN RGG118 \citep{Baldassare15}, and M32 \citep{vandenBosch10}.}
\label{mbh_mgal_pos}
\end{figure}

With our best-fit BH mass and upper limit, we can examine NGC~404 in the context of galaxy scaling relations.  \citet{Kormendy13} and \citet{McConnell13b} have shown that for ETGs, there is a tight correlation between bulge mass and BH mass.  The bulge mass of NGC~404 was determined in S10; with $M_H=-19.62$ and an $M/L_H=0.63$ as determined from spectra, the total bulge mass for NGC~404 is $9.5 \times 10^8$\Msun.  S10 find the S\'ersic index of the bulge to be $n\sim2.5$, and this high S\'ersic index is commonly thought to indicate a ``classical bulge'' formed from merging \citep[i.e., Figure 6 of ][]{Fisher10}. The upper limit on \Mbh~for NGC~404 of $1.5\times10^5$\Msun~clearly lies below the bulge-mass galaxy-mass relation seen in other ETGs (see Figure~\ref{mbh_mgal_pos}).  An increasing number of mostly late-type galaxies have been found to fall below the elliptical bulge mass -- BH mass relation, including the precision maser masses \citep{Greene10,Laesker16}.

Assuming bulge dispersion velocity of $40\pm3$~km s$^{-1}$ \citep{Barth02b}, we find BH masses of 0.8 to 3$\times10^5$\Msun~using the \citet{Kormendy13} relation for all galaxies, largely consistent with our BH mass upper limit.  We also note that the dispersion value measured by \citep{Barth02b} is from a 2$\farcs$0$\times$3$\farcs$7 aperture; kinematic data presented by \citet{Bouchard10} suggests a somewhat lower average dispersion ($\sim$35~km s$^{-1}$) within the bulge effective radius; this would imply an even lower mass BH for NGC~404.

The relation between total galaxy mass and BH mass was recently explored by \citep{Reines15}.  While NGC~404 clearly has an extended disk \citep{Williams10}, the total mass of the galaxy appears to be dominated by the bulge: the 2MASS LGA value for the $M_H=-19.70$ \citep{Jarrett03},~\ml$_H$ = 0.34 and $A_H=0.026$ (S10), and thus the total mass of the galaxy is also $\sim$10$^9$\Msun.  Comparing this to \citet{Reines15}, we find that NGC~404 falls towards the bottom edge of active dwarf galaxies with BH masses estimated from single-epoch measurements of a broad line AGN. These, in turn, are located well below the BH mass galaxy mass relation defined by other quiescent BHs with dynamical mass determinations.

Recently, efforts have been made to incorporate the bulge radius or density in the \Mbh--galaxy scaling relations \citep{Saglia16,Vandenbosch16a}.  In particular, \citet{Saglia16} have found that some of the scatter in the e.g.,~\Mbh--$\sigma$ relation is correlated with the average bulge density and radius.  NGC~404's bulge has an effective radius of 0.675~kpc, and a density within this radius of $\rho_{h}\sim6\times10^8$\Msun~kpc$^{-3}$ (Table~\ref{tab_sersic}).   Using the two-dimensional correlations for all ETGs and classical bulges (CorePowerEClass) from \citet{Saglia16}, the prediction from the \Mbh--$\sigma$--$r_h$ relation is $3.7\times10^5$\Msun, while the prediction from the \Mbh--$\sigma$--$\rho_h$ relation is $4.5\times10^5$\Msun.  Our upper limit is lower than both these predictions by a factor of $\sim$3.

These findings fit in a scenario of co-evolution of BH and classical-bulge masses, where core ellipticals are the product of dry mergers of power-law bulges and power-law Es and bulges the result of (early) gas-rich mergers and of disk galaxies. In contrast, the (secular) growth of BHs is decoupled from the growth of their pseudo bulge hosts, except when (gas) densities are high enough to trigger the feedback mechanism responsible for the existence of the correlations between massive BH and galaxy structural parameters.

With the best fitting BH mass value,~\Mbh~= $7.0\times 10^4$\Msun, its sphere of influence radius is calculated based on the dynamics: $r_g=G$\Mbh$/\sigma^2$, where $\sigma$ is the stellar velocity dispersion of the bulge, $G$ is the gravitational constant. We find $r_g=0.3\pm0.1$~pc or $0\farcs02\pm0\farcs01$. Alternatively, we also estimate the BH sphere of influence radius via our mass model, where $r_g$ is the radius at which the enclosed stellar mass within that radius is equal to twice the black hole mass \citep{Merritt15}. The latter definitions of $r_g$ gives its value of $0.35\pm0.07$~pc or $0\farcs023\pm0\farcs018$. Both methods give a consistent value of the sphere of influence of the BH in NGC~404.  The fact that these values are similar to the pixel size of Gemini/NIFS and WFC3, and smaller than their spatial resolution further justifies the upper limit placed on the BH with our dynamical models.

\subsection{The NGC~404 NSC}\label{ssec:ngc404nsc} 

If we assume the NSC is described by the inner two components of our \textsc{galfit} mass model, the stellar population mass estimate is $\sim$1.35$\times10^7$\Msun~(Table~\ref{tab_sersic}).  From the JAM modeling, we find that the mass scaling factor is $\gamma=0.890_{-0.060}^{+0.045}$, thus, the dynamical mass is $\sim$(1.2$\pm$0.2)$\times$10$^7$\Msun, somewhat lower than this population mass estimate (which assumes a Chabrier IMF). This mass is roughly consistent with the $1.1\times10^7$\Msun~found for the NSC in S10; we note that a different definition of the NSC was used in S10 -- they did not fit the inner portion of the cluster, and the mass was calculated based on the best-fit \ml~ratio multiplied by the luminosity of both the central un-fit component plus the fitted NSC component.  The contrast is clear in comparing the effective radii; the S10 NSC has an effective radius of 0$\farcs$74 (11~pc) while the best-fit here has an effective radius of $\sim$1$\farcs$0 (15~pc), which is estimated from the central two S\'ersic components combined (see Table~\ref{tab_sersic}).  This latter value is on the large side, but within the distribution of NSC sizes \citep[e.g.,][]{Georgiev14}. Our NSC mass estimates are consistent with the prediction of $M_{\rm NSC}-M_{\rm Gal,dyn}$ scaling relationship of \citet{Scott13a} and \citet{Ferrarese06}.

As was found in S10, the dynamical and stellar population mass estimates for the NGC~404 nucleus agree quite well, with a mass scaling factor $\gamma=0.890$.  This provides some evidence that the IMF in the nucleus of NGC~404 is roughly consistent with the assumed Chabrier IMF.  This relatively ``light'' IMF is consistent with the ratios of dynamical masses to stellar population model estimates in massive globular clusters in M31 \citep{Strader11b}, and in massive ultra-compact dwarf galaxies (UCDs) that are likely stripped nuclei \citep[][Ahn et al. \emph{in prep}]{Mieske13, Seth14}. There is growing evidence that the IMF is not universal among galaxies \citep[e.g.,][]{Cappellari12, Conroy12, Kalirai13, Cappellari13b, Spiniello14}; given the high stellar density of the NGC~404 nucleus, this suggests that a parameter other than star formation rate density may be driving these IMF variations.

Finally, we note again that the STIS data provides strong evidence for the dominance of a $\sim$1~Gyr old stellar population in the inner few parsecs of the NGC~404 NSC.  Compared to the S10 stellar populations determined from a ground-based spectrum of the entire NSC, this 1~Gyr population is clearly enhanced in the center.  S10 found that the cluster counter-rotates between this inner portion and the outer portion of the NSC; the new results suggest that the counter-rotating component may be a 1~Gyr old addition to the pre-existing NSC.

\section{Conclusions}\label{sec:concl}

We present a new analysis of the nucleus of NGC~404 with data from~\hst/WFC3 and STIS.  We use this data to analyze the stellar populations of the nucleus and combine this with kinematic data from Gemini/NIFS to constrain the mass of the BH and NSC at the center of NGC~404.

\begin{enumerate}

\item We develop a method to incorporate variations in the stellar~\ml~into the dynamical modeling to constrain the BH mass in NGC~404.  Specifically, we use spectroscopically determined~\ml~spatial variations of the nuclear region and use these to create~\ml~vs. color relations appropriate to the local stellar populations present in the nucleus.  We then use these relations to create a mass map of the nucleus.  The spatial variabilities in~\ml are thus directly incorporated into our dynamical model. Incorporating stellar population-based models are critical for getting good dynamical constraints on the lowest mass BHs, as most of these are located in NSCs with complicated and spatially varying stellar populations.

\item Our derived color--\ml~relation is inconsistent with previously published relations.   It is steeper and much lighter than the relation from \citet{Bell03}, while it is shallower than the color--\ml~relations of \citet{Roediger15}.  This suggests that creating a color--\ml~relation based on local populations (and not based on a library of SFHs), can result in significant differences in the inferred mass profile of a galaxy.

\item Jeans anisotropic models of the stellar kinematics with the derived mass map gives a BH mass of \Mbh~$=7.0^{+2.0}_{-0.6}\times10^4$\Msun, a mass scaling factor (the ratio of the dynamical to stellar population mass) of $\gamma=0.890_{-0.060}^{+0.045}$, an anisotropy parameter $\beta_z=0.05^{+0.05}_{-0.15}$, and an inclination angle $i=20.5^{+1.0}_{-2.0}$.  The BH mass is consistent with zero at the 3$\sigma$ level, thus, we present a 3$\sigma$ upper limit to the mass of $1.5\times10^5$\Msun.  This BH mass upper limit suggests NGC~404 falls clearly below scaling relations between the BH mass and the bulge or total mass while it is consistent with the $M-\sigma$ relation.

\item We find $\sim$20\% variability at the center of NGC~404 in three filters (\FB, \FV, \FI) over a 15 years period.  This variability, combined with previous, less robust claims of variability \citep{Maoz05, Seth10}, provides strong evidence for an accreting BH at the center of NGC~404.   Furthermore, the STIS spectra are best fit including a 17\% AGN continuum fraction in the central most pixels.  Thus, we add significant strength to the previous claims of an AGN at the center of NGC~404, and clearly locate the AGN at the center of the NSC.  This makes NGC~404 the lowest mass central BH with a dynamical constraint.

\item  Population synthesis fits of the STIS spectra show that the central portion of the NSC at radii $\lesssim$0$\farcs$3, $\sim$75\% of the stars have ages of 1~Gyr; this is significantly higher than is found in the cluster as a whole and provides further evidence that the central counter-rotating component of the NSC found by S10 was formed in a minor merger.

\item The dynamical mass of the NSC is found to be $(1.2\pm0.2)\times10^7$\Msun.  The ratio of the dynamical mass estimate to the population mass estimates based on spectral synthesis is $\gamma=0.890_{-0.060}^{+0.045}$.   This suggests the NSC has an IMF that is similar to the Chabrier IMF assumed in the stellar population mass estimate.

\end{enumerate}
\acknowledgments

The authors would like to thank Yiping Shu and Antonio Montero-Dorta at the University of Utah, Physics and Astronomy Department for their helpful discussions about programming, debugging, and running FSPS model for SSPs; Joel Roediger for sharing his data with us; and the University of Utah, Physics and Astronomy Department, for supporting this work. DN and ACS acknowledge financial support from~\hst~grant GO-12611 and from NSF grant AST-1350389. MC acknowledges support from a Royal Society University Research Fellowship. Research by A.J.B.~is supported in part by NSF grant AST-1412693.


\bibliographystyle{apj}
\bibliography{ngc404}

\end{document}